\documentclass{elsarticle}
\usepackage{url}

\usepackage{natbib}
\usepackage{subfigure}
\usepackage{pifont}
\usepackage{geometry}
\usepackage{graphicx}
\usepackage{txfonts}
\newcommand{\farcs}{\mbox{\ensuremath{.\!\!^{\prime\prime}}}}% 

\newcommand{\slv}{{\sl v}}      % use instead of $v$ in text mode
\newcommand{\msv}{{\mathsf v}}  % use in math mode

\newcommand{\civ}{C\,{\sc iv}}
\newcommand{\ebv}{$E$$($$B$$-$$V$$)$}

\newcommand{\hi}{H\,{\sc i}}
\newcommand{\hii}{H\,{\sc ii}}

\newcommand{\kms}{km\,s$^{-1}$}

\newcommand{\alii}{Al\,{\sc ii}}

\newcommand{\aliii}{Al\,{\sc iii}}

\newcommand{\cai}{Ca\,{\sc i}}
\newcommand{\caii}{Ca\,{\sc ii}}

\newcommand{\feii}{Fe\,{\sc ii}}
\newcommand{\feiii}{Fe\,{\sc iii}}
\newcommand{\feiv}{Fe\,{\sc iv}}

\newcommand{\mgi}{Mg\,{\sc i}}
\newcommand{\mgii}{Mg\,{\sc ii}}

\newcommand{\Nv}{N\,{\sc v}}

\newcommand{\SIiv}{Si\,{\sc iv}}

\newcommand{\nobal}{SDSS~J0259$+$0048}	
\newcommand{\otbal}{SDSS~J0300$+$0048}	
\newcommand{\febal}{SDSS~J2215$-$0045}	
\newcommand{\aox}{$\alpha_{ox}$}
\newcommand{\alox}{\alpha_{ox}}
\newcommand{\aloxUV}{\alpha_{ox}(UV)}
\newcommand{\fuv}{$f_{2500}$}
\newcommand{\luv}{$l_{2500}$}
\newcommand{\lx}{$l_{2\,keV}$}
\newcommand{\fx}{$f_{2\,keV}$}

\newcommand{\aj}{Astron.~J. }%
\newcommand{\apj}{Astrophys.~J. }%
\newcommand{\apjl}{Astrophys. J. Lett. }%
\newcommand{\apjs}{Astrophys. J. Suppl. Ser. }%
\newcommand{\aap}{Astron. Astrophys. }%
\newcommand{\mnras}{Mon. Not. R. Astron. Soc. }%
\begin{document}

% 44.998667   0.803778
% 45.002333   0.807778
% 333.799750 -0.763861

\title{{\em Chandra} X-ray Observations of Two Unusual BAL Quasars} 

\author[1]{Jesse A. Rogerson}
\ead{rogerson@yorku.ca}

\author[1]{Patrick B.Hall}
\ead{phall@yorku.ca}

\author[2]{Stephanie A. Snedden}

\author[3]{Michael S. Brotherton}

\author[4]{Scott F. Anderson}

\address[1]{Department of Physics and Astronomy,
York University, Toronto, Ontario M3J 1P3, Canada}
\address[2]{Apache Point Observatory, P.O. Box 59, Sunspot, NM 88349-0059}
\address[3]{Department of Physics and Astronomy, University of Wyoming, 
Laramie, WY 82071}
\address[4]{University of Washington, Department of Astronomy, Seattle, WA
98195}

\begin{abstract}
We report sensitive {\em Chandra} X-ray non-detections of two unusual,
luminous Iron Low-Ionization Broad Absorption Line Quasars (FeLoBALs).
The observations do detect a non-BAL, wide-binary companion quasar to one
of the FeLoBAL quasars.  
We combine X-ray-derived column density lower limits (assuming solar metallicity)
with column densities measured from ultraviolet spectra
and CLOUDY photoionization simulations to explore whether constant-density
slabs at broad-line region densities can match the physical parameters 
of these two BAL outflows, and find that they cannot.
In the ``overlapping-trough'' object \otbal, 
we measure the column density of the X-ray absorbing gas to be 
$N_H \geq 1.8 \times 10^{24}$ cm$^{-2}$.  From the presence of \feii\ UV78
absorption but lack of \feii\ UV195/UV196 absorption, we infer the density
in that part of the absorbing region to be $n_e \simeq 10^{6}$ cm$^{-3}$.  
We do find that a slab of gas at that 
density might be able to explain this object's absorption.
In the \feiii-dominant object \febal, the X-ray absorbing column density of
$N_H \geq 3.4 \times 10^{24}$ cm$^{-2}$ is consistent with the \feiii-derived
$N_H \geq 2 \times 10^{22}$ cm$^{-2}$ provided the ionization parameter is
$\log U>1.0$ for both the $n_e=10^{11}$ cm$^{-3}$ and $n_e=10^{12}$ cm$^{-3}$ 
scenarios considered (such
densities are required to produce \feiii\ absorption without \feii\ absorption).
However, the velocity width of the absorption rules out its being concentrated
in a single slab at these densities.  Instead, this object's spectrum can be
explained by a low density, high ionization and high temperature disk wind that
encounters and ablates higher density, lower ionization \feiii-emitting clumps.
\end{abstract}


\maketitle

%\clearpage

\section{Introduction}
\label{introduction}

Broad Absorption Line (BAL) quasar spectra are not as common as typical 
quasar spectra but provide a unique look at the central regions of Active 
Galactic Nuclei (AGN).  A BAL quasar is characterized by absorption troughs 
from gas with blueshifted outflow velocities of typically 10\%\ the speed of 
light (0.1$c$) \citep{wea91}.
The lower limit on the width of a BAL trough is in part a matter of definition;
the traditional required minimum width is 2000 km s$^{-1}$ \citep{wea91}, but a
minimum width of 1000 km s$^{-1}$ has also been used \citep{trump06}.  Such 
velocity widths are larger than essentially all galactic wind outflows, and 
thus ensure a sample of outflows driven predominantly by AGN; of course, 
narrower AGN-driven outflows can and do exist.

BAL quasars themselves are distributed into three subtypes:
high-ionization (HiBAL), low-ionization (LoBAL), and iron LoBAL (FeLoBAL).
HiBAL quasars show absorption from only
relatively high-ionization species such as \civ, \Nv\ and \SIiv.  LoBAL quasars
also show absorption from low-ionization species such as \mgii, \aliii\
and \alii.  FeLoBAL quasars are LoBAL quasars with
absorption from one or more excited states of \feii\ or \feiii.
Note that a range of ionization stages is seen even in HiBAL
quasars, and that in LoBAL and FeLoBAL quasars the ionization simply extends to
lower ionization stages \citep{sdss123}.

Approximately 10\% of quasars in optically selected samples with spectra
covering rest-frame 1400-1550\,\AA\ exhibit a trough or troughs
which can be considered broad absorption.  The true fraction is
expected to be higher due to selection effects which bias surveys against the
detection of BAL quasars.  \citet{hf03} reported a corrected traditional BAL
quasar fraction of $22\pm 4$\% using their sample of 42 bright ($B_J<19$) BAL
quasars.  \cite{sdssbal} estimated a corrected traditional BAL quasar fraction
of $15.9\pm 1.4$\% using a sample of 224 BAL quasars with $i\lesssim20$.
\citet{trump06} report an uncorrected traditional BAL quasar fraction of $10.4
\pm 0.2$\% using a sample of 1756 BAL quasars with $i\lesssim 20$, or an
uncorrected fraction of $26.0\pm 0.3$\% using a less conservative
``absorption index'' criterion to define a BAL quasar
(uncertainties from Poisson statistics only).  By comparing
the \citet{trump06} catalog to the near-infrared 2MASS database, \cite{dss08}
report a corrected BAL fraction of $43\pm 2$\% for luminous quasars ($M_{K_s}
\lesssim -30.1$) using the absorption index criterion.
However, \cite{ksgc08} find that the absorption index BAL quasar criterion
includes a large number of false positives.  Taking that into account, they
find uncorrected and corrected BAL fractions of 13.6\% and 17$\pm$3\%.
Scaling the results of \cite{dss08} to account for the contamination found by
\cite{ksgc08} yields a BAL fraction of 23\%, which is also the upper limit
BAL fraction quoted by \cite{ksgc08}.
Lastly, \cite{g07} find an uncorrected BAL fraction of 11\%,
corrected to 23\% in \cite{gb08}.  In summary, there is general agreement that $\simeq$23\% of quasars exhibit
BAL troughs after selection effects are taken into account.

The fraction of quasars exhibiting BAL troughs could be due to an orientation
effect, such that all or most quasars have BAL outflows covering some of the
lines of sight along which they would be seen as quasars (e.g.,
\citealt{tur86}).  Alternatively, it could be due to an evolutionary effect,
such that all or most quasars have BAL outflows for some fraction of their
lifetime (e.g., \citealt{sh87}).  Of course, some combination of the above
is also possible (e.g., \citealt{mor88}).  In any case, these outflows may
have mass loss rates comparable to the accretion rates required to power
quasars \citep{SKC05}.  Thus, understanding BAL quasars and BAL outflows
is important for understanding quasars as a whole.  

Observations show that BAL quasars are X-ray weak compared with regular quasars (e.g. \citealt{GSA95}, \citealt{gea06}, and references therein).
The disk wind model of \citet{mcgv} explains this and other properties of BAL quasars well.
In that model, the X-ray weakness is attributed to intrinsic absorption from 'hitch-hiking' gas between the BAL outflow and the X-ray emitting region.  The nearly completely ionized metal atoms in the hitch-hiking gas
are kept ionized by capturing electrons
and then immediately absorbing X-rays from the quasar, just as
an \hii\ region is kept ionized by protons capturing electrons
and then immediately absorbing photons of energies 13.6 eV or greater.
Because highly ionized metal atoms can only efficiently absorb X-rays,
the hitch-hiking gas protects the BAL outflow from overionization by X-rays
while transmitting the ultraviolet (UV) radiation that
accelerates the BAL outflow to high velocities.\footnotemark[1]  We assume a BAL quasar has an intrinsically normal X-ray spectrum but is shielded by some absorbing gas, so that constraints can be placed on the hydrogen column density required to absorb the X-ray flux.

\footnotetext[1]{The hitch-hiking gas may also act as an X-ray reflector and a soft X-ray emitter, but we do not consider those effects herein.}

It is also possible that some or all BAL quasars are intrinsically X-ray weak.  It has been shown by \citet{GCV08} that some BAL quasars may be less X-ray absorbed than previously thought, and the X-ray weakness is due to a differing spectrum from typical quasars.  The authors admit there may be some factors that could affect this result.  These include the X-ray energy range used for spectral analysis, partial covering of the source by the X-ray absorber, or an ionized X-ray absorber.  \citet{SBCG10} go on to show that when an ionized absorber is considered, the column density of intrinsic absorption significantly increases.  Although the possibility of intrinsic X-ray weakness cannot be fully rejected by current data sets, traditional BAL quasars (with BI$>0$, where BI is the BALnicity index) appear to be always X-ray absorbed \citep{SBCG10}.  Both the above studies use some BAL quasars from \citet{trump06}, which may not be actual BAL quasars \citep{ksgc08}.

%{\bf Absorption in the rest-fram UV is dependent on the column density of relatively low-ionization absorbers, whereas the X-ray absorption depends on the column density of relatively high-ionization absorption.  As a result, a high- and low-ionization absorbers of the same column density could result in vastly different X-ray and UV absorption properties.}

Recent sky surveys have also uncovered a slew of BAL quasars which exhibit 
unusual properties as compared to typical BAL quasars (e.g., \citealt{sdss123,
pecbal,duc02}).  Investigation of these properties is as worthwhile as the study of BAL
quasars themselves, as the most unusual objects will define the parameter space
spanned by BAL outflows in general.  If we are to have a complete understanding
of quasars, our models must explain all normal and unusual quasar 
behaviour.

\begin{figure}
\centering

\subfigure{
\includegraphics[scale=0.6]{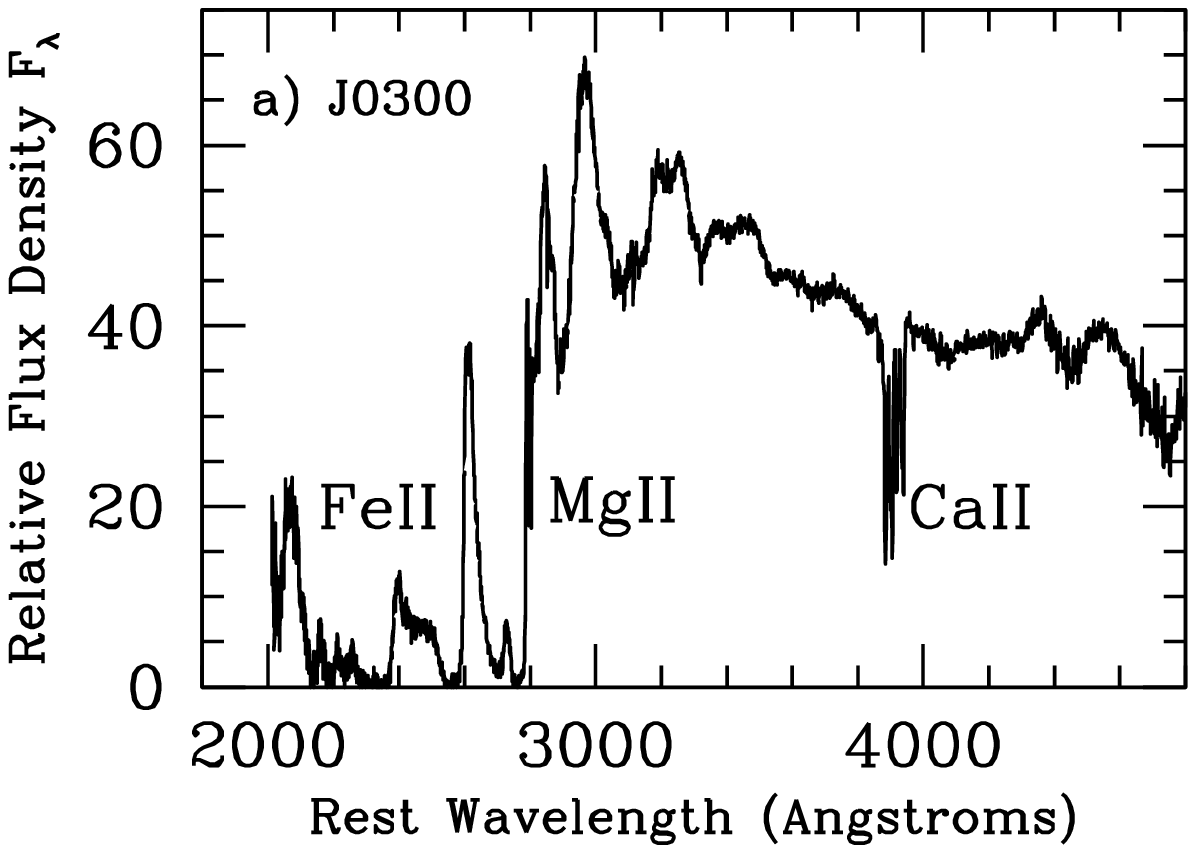}
}
\subfigure{
\includegraphics[scale=0.6]{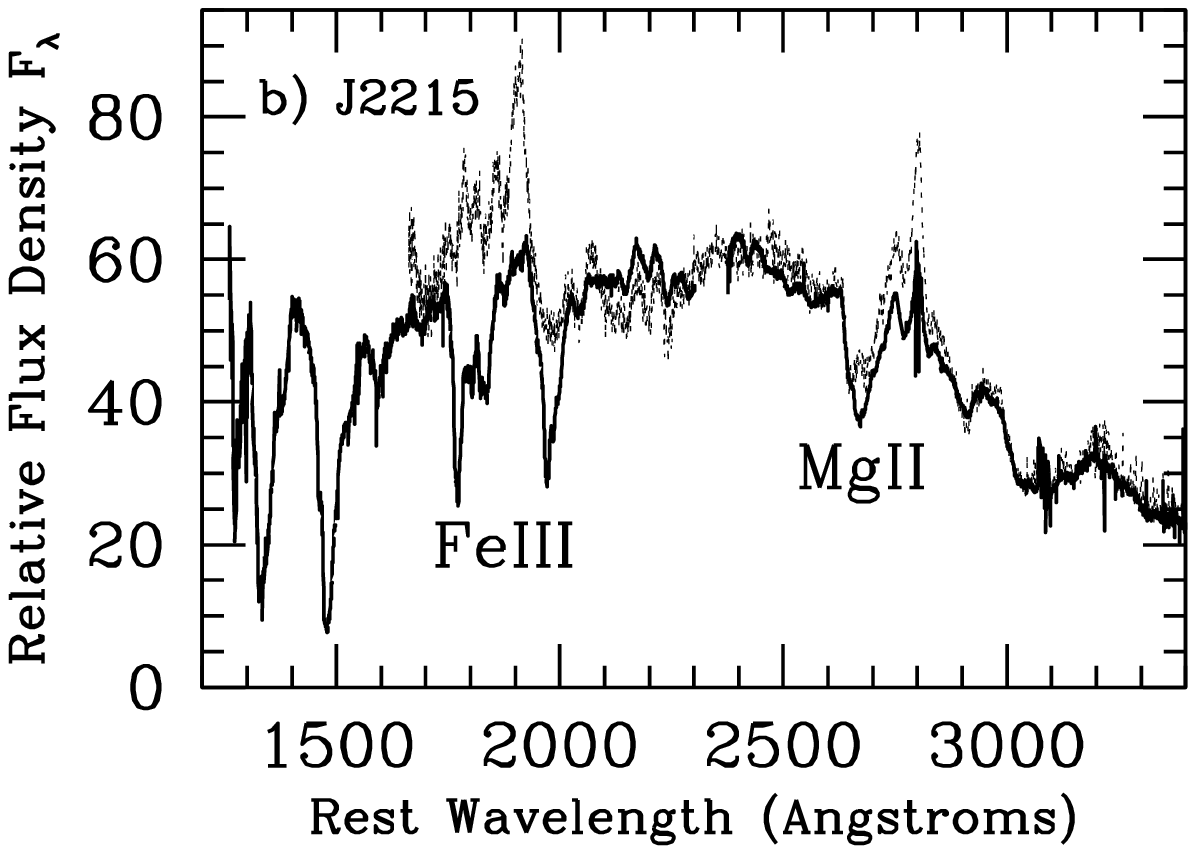}
}
\caption{Spectra of the two target quasars, with selected absorption troughs
marked.  The vertical axes give $F_\lambda$ in units of
10$^{-17}$ erg s$^{-1}$ cm$^{-2}$ \AA$^{-1}$.
{\bf a.} The overlapping-trough object SDSS J0300+0048.
Below 2800\,\AA, \feii\ and other troughs overlap
to absorb nearly all the flux from the quasar.
{\bf b.} The \feiii-dominant BAL quasar SDSS J2215$-$0045 (solid) and a partial
estimate of its intrinsic, unabsorbed spectrum (dashed), from \citet{sdss199}.
The UV34 and UV48 multiplets of \feiii\ are very strong in absorption,
while only upper limits can be put on absorption from any \feii\ multiplet.
}
\label{spectra}
\end{figure}

In this paper we discuss continued research into two such unusual BAL quasars.
SDSS J030000.56+ 004828.0 (hereafter \otbal, Fig. \ref{spectra}a) is an ``overlapping-trough'' BAL
quasar at $z=0.89191$ with nearly complete UV absorption below rest-frame 2800~\AA\ 
(\citealt{sdss123}; \citealt{sb2}).
% used high-resolution spectroscopy to measure the \caii\ column density in this object.  
%The \caii \lala3934,3969, \mgii \lala2796,2803, and \mgi \lam2852 column densities are 
%the largest reported to date for any BAL outflow.  
This FeLoBAL 
quasar has a non-BAL binary companion, SDSS J025959.68$+$004813.6 (hereafter SDSS J0259+0048), which is located 19\farcs5 away from \otbal\ at a redshift 
$z=0.894$ ($\Delta \msv=330 \pm 160$\,\kms).

The second object is SDSS J221511.94$-$004549.9 (hereafter \febal, Fig. \ref{spectra}b), a 
reddened FeLoBAL with detached \mgii, \aliii, 
\alii, and \feiii\ UV34,48 absorption \citep{sdss123}.  This quasar is at $z=1.4755$ as 
calculated by the associated \mgii\ absorption. 
%All other initial observations of this object are summarized in \citet{sdss123}.  
\febal\ has 
absorption from \feiii\ but there are no absorption features from \feii, making 
it a very rare and unusual find.  \citet{sdss199} discuss how the work of 
\citet{dek02b} shows that the presence of \feiii\ but lack of \feii\ absorption
restricts the outflow to high densities and a narrow range of column densities.

X-ray data has been obtained using the Chandra X-ray Observatory and 
analyzed to help understand these two unusual BAL quasars.  
We discuss our data and methods in \S \ref{Xray}, present the results
of our observations in \S \ref{otbal&nobal} and \S \ref{febal},
discuss their implications in \S \ref{discuss}
and summarize our results in \S \ref{conclusion}.  
We adopt the cosmology of \cite{wmap3}:
H$_0$$=73.2$, $\Omega_M=0.259$, and $\Omega_\Lambda=0.741$.

%\clearpage

\section{X-ray Data and Analysis}
\label{Xray}

Observations of our two targets were carried out in Chandra Cycle 4 
using the Advanced CCD Imaging Spectrometer (ACIS).

We use the observed-frame energy range 0.5-6.65 keV: 0.5 keV is an obvious lower energy cutoff due to the 
increased background (and higher throughput degradation) at low energies,
and 6.65 keV was chosen as the upper limit because it has the same effective 
area as the 0.5 keV lower limit and because the effective observing area drops
quickly at energies $>$6.65 keV. The source count rate in the energy range
0.5-6.65 keV is expected to be a factor of 100 higher than in the range
6.65-13 keV, while the expected background count rate is the same
in these energy ranges.\footnotemark[2]

\footnotetext[2]{See http://cxc.harvard.edu/proposer/POG/html/index.html}

Given the small number of detections of X-ray photons over a finite time period,
we are limited by small-number statistics rather than by the background level.
%There are primarily two methods for handling small-number statistics, both 
%of which are based on Poisson statistics.
To determine upper limits at specified confidence intervals in the face of
small-number statistics, we use the Bayesian method of \citet{kbn91}
rather than the frequentist method outlined in \citet{geh86}.
The Bayesian method is favored due to its ability to work with 
non-zero background flux when in the zero source count regime.

\subsection{AGN X-ray-UV Luminosity Correlation}
\label{method}

The X-ray and UV luminosities of regular (non-BAL) quasars are observed to have a strong correlation across multiple decades of UV luminosity; this offers a useful tool for measuring the X-ray weakness of a BAL quasar, which tend to have far weaker X-ray luminosity than a typical quasar.  The correlation is usually expressed in terms of the logarithm of the ratio between the 2 keV (\lx) and 2500\AA\ (\luv) rest-frame specific luminosities, denoted as $\alpha_{ox}$:
\begin{equation}
\alpha_{ox} = 0.372 \log (l_{\rm 2\,keV}/l_{\rm 2500})
\label{alphaoxobs}
\end{equation}
This correlation has been quantified carefully in a recent study by \citet{s06} by observing the \lx\ and \luv\ of 333 AGNs (from multiple surveys).  The best fit to the data is:
\begin{equation}
\log(l_{2\,keV}) = (0.721 \pm 0.011)\log(l_{2500}) + (4.531 \pm 0.688)
\label{xraylum}
\end{equation}
Thus, provided a UV luminosity, the expected X-ray luminosity can be calculated.  For typical quasars, \aox\ can be written as a function of \luv, as in \citet{s06}:
\begin{equation}
\alpha_{ox}(UV) = (-0.137 \pm 0.008)\log(l_{2500}) + (2.638 \pm 0.240)
\label{alphaoxexp}
\end{equation}

To characterize the X-ray weakness we compare the observed \aox\ value, calculated using the observed \luv\ and \lx\ in equation (\ref{alphaoxobs}), to the expected \aox\ value $\aloxUV$.  The difference between the two is denoted $\Delta \alox$:
\begin{equation}
\Delta\alpha_{ox} = \alpha_{ox} - \alpha_{ox}(UV)
\label{deltaalphaox}
\end{equation}

By exploiting the strong correlation between \luv\ and \lx\ in ordinary quasars, we can quantify the X-ray weakness of our targets.

To determine \luv\ for our targets requires estimating \fuv\ for their
underlying continua.  We used the spectra available for each object in SDSS
Data Release Six \citep{dr6}.
We used either the smoothed or extrapolated flux at 2500 \AA\ rest-frame as 
\fuv; details are given in the individual objects' discussions below.
We have assumed a value of 10\%\ as the uncertainty on the UV flux for the following discussion.
The luminosity distance to each quasar was calculated following \citet{pen99}
based on the quasar's redshift and our adopted cosmology.

%\clearpage

\section{Results: SDSS J0300+0048 and SDSS J0259+0048}
\label{otbal&nobal}
%--tstop-tstart = 6743.47529

\otbal\ and \nobal\ %(with redshifts of 0.89191 and 0.894 respectively) 
were observed simultaneously for 6743.5 seconds on December 19, 2002 (UT).
No X-ray photons were detected within a 2\farcs5 radius of the position of 
\otbal.  \nobal\ was detected, with 39 photons within 2\farcs5 of its position.
The measured background of the image is such that we expect only 
$0.244 \pm 0.014$ observed-frame 0.5-6.65 keV background photons in the detection apertures.  

%-----------------------------NO BAL------------------------------------------%
%\clearpage

\subsection{Results: SDSS J0259+0048} %0.894
\label{nobal}
We first analyze the \nobal\ results, as it is a non-BAL quasar and is thus 
expected to be unremarkable.  As a test of our methods, we seek to reproduce 
the observed X-ray photon count of 39, by using only the observed UV flux and the relations in \S \ref{method}.

%To determine \luv\ for our targets requires estimating \fuv\ for their 
%underlying continua.  We used the spectra available for each object in SDSS 
%Data Release Six \citep{dr6}.  For \nobal\ we averaged the two available 
%spectra.  We then smoothed the spectrum and adopted the resulting flux value at
%2500 \AA\ rest-frame as \fuv, ignoring a $\sim$10\% correction for \feii\ 
%emission in this object at that wavelength. The resulting value is \fuv\ 
%$=(0.55 \pm 0.05) \times 10^{-27}$ erg s$^{-1}$ cm$^{-2}$ Hz$^{-1}$ for 
%\nobal\ at $z=0.894$.
%
For this quasar at $z=0.894$ we find
\fuv$=(0.55 \pm 0.05) \times 10^{-27}$ erg s$^{-1}$ cm$^{-2}$ Hz$^{-1}$
and \luv\ $=(2.00 \pm 0.20) \times 10^{30}$ erg s$^{-1}$ Hz$^{-1}$.
%
%The luminosity distance to the quasar is calculated following \citet{pen99} 
%based on the redshift and adopted cosmology; thus, we find \luv\ $=(2.00 \pm 
%0.20) \times 10^{30}$ erg s$^{-1}$ Hz$^{-1}$.  

Using equation (\ref{xraylum}) we therefore expect $\log$(\fx) $=-31.2^{+0.8}_{-0.7}$ erg s$^{-1}$cm$^{-2}$ Hz$^{-1}$. Given the predicted X-ray flux, we expect 
%$2.04^{+0.77}_{-0.76}$ counts in the log in our image (via webPIMMS\footnotemark[2]).  That is 
2.8 times the observed count rate of 1.59 in the log, but the difference is $<1
\sigma$.  Thus, our method accurately reproduces the observed X-ray photon 
count of an unremarkable quasar.

\footnotetext[3]{http://cxc.harvard.edu/toolkit/pimms.jsp: the online version of Portable, Interactive Multi-Mission Simulator, providing count-rate estimations and predictions for {\em Chandra}.}

Using equation (\ref{alphaoxobs}) we calculated an observed $\alox=-1.63 \pm
0.04$ for \nobal.  The expected \aox\ is calculated by equation 
(\ref{alphaoxexp}).  Plugging in the observed \luv\, we expect $\aloxUV=-1.51 \pm 0.34$.  Thus $\Delta\alox=-0.12\pm0.34$.  Therefore, within the uncertainties, the observed and expected values for \aox\ (and X-ray counts) are the same.

%-------------------------Over Lapping Trough BAL-----------------------------%

%\clearpage

\subsection{Results: SDSS J0300+0048}  %0.89184
\label{otbal}

Using the same method as above, we analyze the observed and predicted X-ray 
quantities for the FeLoBAL quasar \otbal\ at $z=0.89194$ to determine the column 
density of the intervening matter.

The ``detection'' of zero photons in the image results in an upper limit
of 3.00 photons at 95\% confidence \citep{kbn91}, equivalent to
a count rate of $\leq 4.445\times 10^{-4}$ photons s$^{-1}$ in the 
observed bandpass.  This corresponds to an observed X-ray flux upper limit of 
\fx\ $\le 1.793 \times 10^{-33}$ erg s$^{-1}$ cm$^{-2}$ Hz$^{-1}$ at 95\% 
confidence.

To find \fuv\ we fit a power law in $f_\nu$ to 
narrow normalization windows centered at 3060~\AA\ and 4735~\AA\ rest-frame.  
The resulting flux at 2500~\AA\ is \fuv\ $=(3.77 \pm 0.38) \times 10^{-27} $ 
erg s$^{-1}$ cm$^{-2}$ Hz$^{-1}$.  %for \otbal\ at $z=0.89184$.

Equation (\ref{xraylum}) then predicts the X-ray flux of \otbal\ to
be $\log$(\fx) $=-30.58^{+0.77}_{-0.79}$ ergs s$^{-1}$ cm$^{-2}$ Hz$^{-1}$, 
which includes the uncertainty in the power law slope in the UV band.  
Therefore, we expect to see $2.6^{+0.8}_{-0.7}$ X-ray counts in the log in the 
image.  %(webPIMMS).  
This is higher than the observed 95\% confidence limit of $0.5$ in the log.

We characterize this X-ray weakness as follows.  The expected 
\aox\ (from equation \ref{alphaoxexp}) is $\aloxUV=-1.64 \pm 0.35$ and the 
observed \aox\ (from equation \ref{alphaoxobs}) is $\alox \le -2.35$ at 95\%\ confidence.
Therefore the deviation of the observed \aox\ from the predicted value is 
$\Delta \alox \le -0.72$, which is a reduction in \fx\ by a 
factor of 164 from the expected flux.  This reduction in X-ray flux is 
attributed to shielding gas near the quasar, which must have a 
hydrogen column density of $N_H \ge 1.8 \times 10^{24}$ cm$^{-2}$ at the 
quasar redshift, assuming solar metallicity.  
That lower limit was calculated assuming neutral gas, but applies to neutral
or ionized gas.  If the gas is partly or fully ionized,
it will absorb fewer photons per unit column density, and therefore the
required minimum column density will be larger than the above lower limit.

\citet{sdss5BAL} report $\alox \le -2.33$ and $\Delta \alox \le -0.70$ for this object in their summary of BAL quasars from SDSS data release 5.  The values are slightly different from values quoted in this paper because slightly different formulae were used.  Nevertheless, the values are consistent within the uncertainties.

%--------------------------END Calculations-----------------------------------%
%Taken from the Hall et al. (2003) paper - high res spectroscopy...
We can compare our result with the column densities required to explain the UV
absorption, in particular the \caii\ column density $N_{CaII}=(7.13 \pm 1.15)
\times 10^{14}$ cm$^{-2}$ \citep{sb2}. Because its ionization potential is less
than that of \hi, \caii\ becomes the dominant calcium ion only at large columns
behind an \hi\ ionization front (\S 5.2 of \citealt{sb2}).  \cite{fp89} found 
that for clouds of density $10^{9.5}$ cm$^{-3}$, \caii\ is dominant only at a 
column densities $N_H \ge 6.3 \times 10^{24}$ cm$^{-2}$, more than a factor 
of ten higher than the column density of the \hi\ ionization front.

We have used the photoionization simulator CLOUDY\footnotemark[4] 
to investigate the absorption in various ions for
a representative Broad Line Region (BLR) (density of $n_e=10^{11}$ cm$^{-3}$, 
$\log U=-1.5$, where $U$ is the ionization parameter) \citep{blr03}.  
%CLOUDY\footnotemark[3] is a photoionization simulator that uses numerical techniques to simulate an environment where a host of microphysical effects and parameters must be taken into account (e.g. density, kinetic temperature, ionization, radiative transfer, etc.), rendering analytical solutions nearly impossible.  
Note that our CLOUDY simulations were run with solar metallicity 
and no dust obscuration.  Even at $N_H = 10^{25}$ 
cm$^{-2}$, \caii\ is not the dominant calcium ion.  However, the ionic fraction
of \caii\ immediately behind the \hi\ ionization front is $\sim$1\% for 
$n_e=10^{11}$ cm$^{-3}$, significantly higher than the $\sim$0.3\% seen for 
$n_e=10^{9.5}$ cm$^{-3}$. Given that the abundance of \caii\ is $-5.64$ in the 
log, each $N_H=10^{22}$ cm$^{-2}$ behind the \hi\ ionization front yields 
$10^{(22-2-5.64)}$ worth of \caii: $N_{CaII}=2.29 \times 10^{14}$ cm$^{-2}$.  
\cai\ has a similar behavior to \caii\, though its ionic abundance immediately 
behind the \hi\ ionization front is only 0.0001\%.

\footnotetext[4]{http://www.ferland.org/cloudy/}

Thus, to match the \caii\ column observed in \otbal\ with standard BLR 
parameters requires only $N_H=3.11 \times 10^{22}$ 
cm$^{-2}$ behind the hydrogen ionization front.  This agrees with the 
constraint on the total $N_H$ behind the front from the observed upper limit on
the \cai\ column density: $N_H \le 7 \times 10^{23}$ cm$^{-2}$.
%-----------------------------------------------------------------------------%

\begin{figure}

\subfigure{
\includegraphics[scale=0.35]{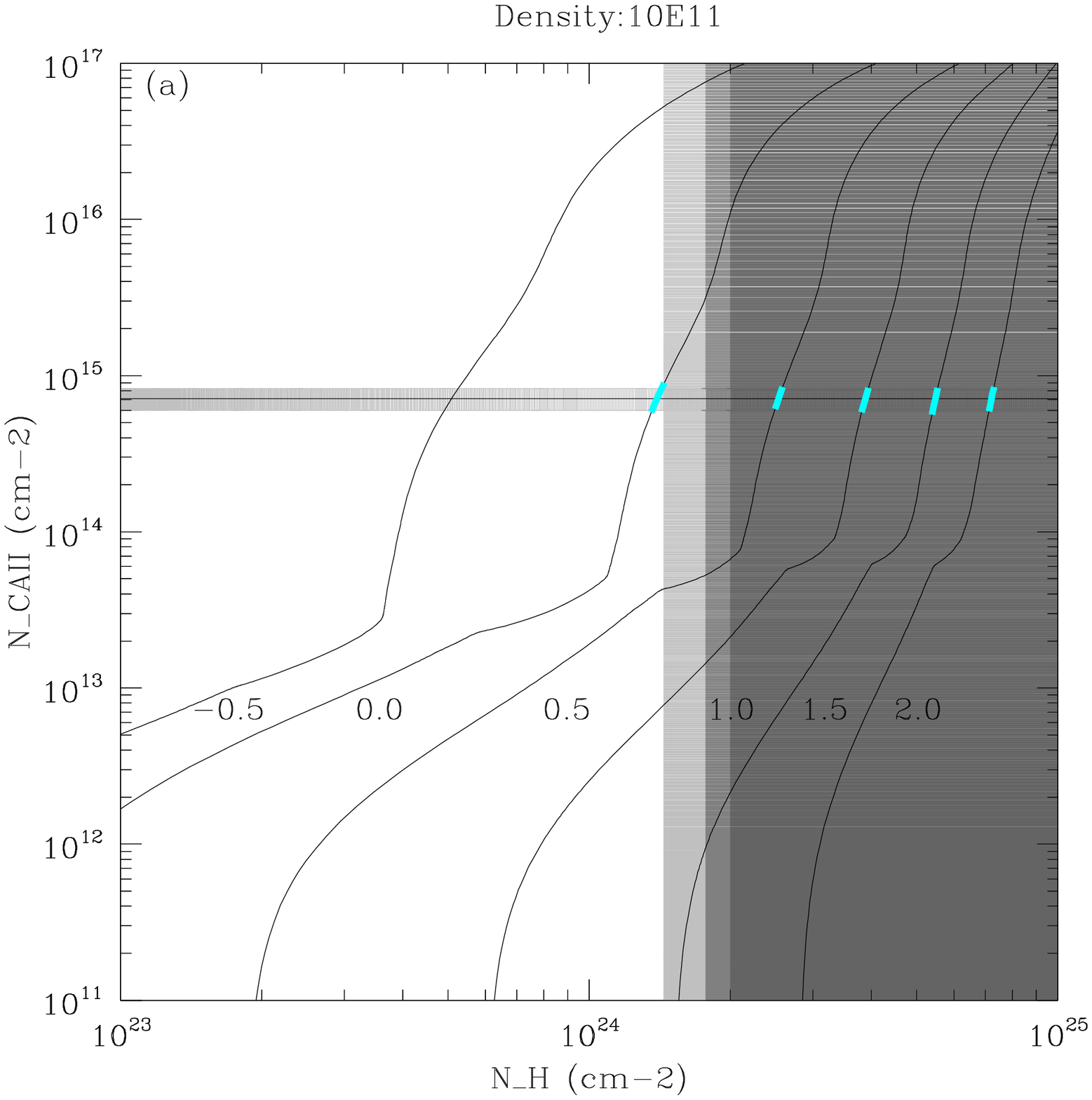}
}
\subfigure{
\includegraphics[scale=0.35]{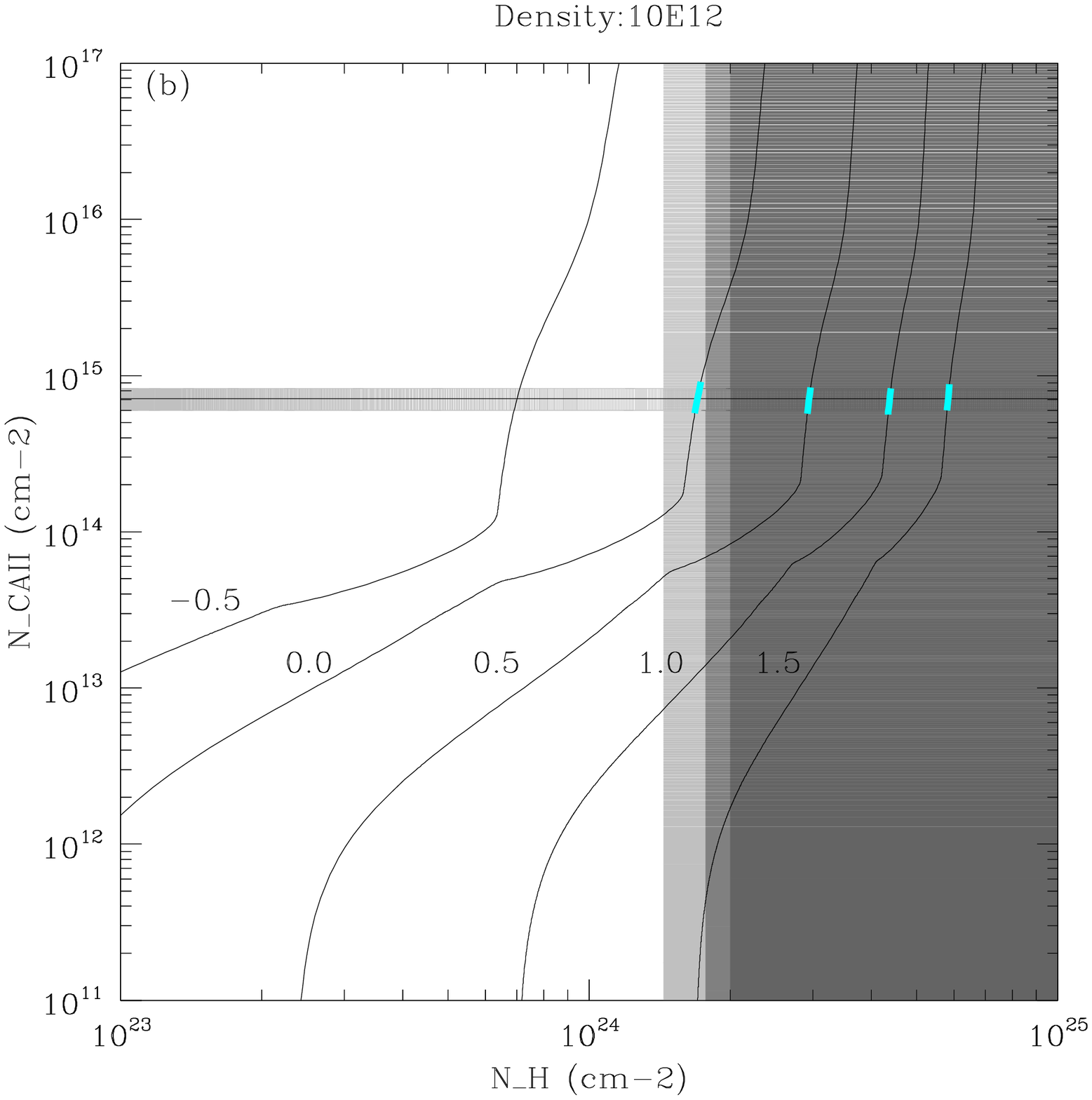}
}
\caption{The behaviour of the \caii\ column density for varying ionization
parameter and density. Shaded regions represent constraints posed by column 
densities of \caii\ and hydrogen in \otbal.
The vertical lines separating the dark, 
light, and lightest shades of grey are the lower limits for the X-ray 
calculated hydrogen column density for the 90\%, 95\% and 99\% confidence 
intervals, respectively.  The light grey horizontal bar represents \caii\ 
column density and its uncertainty.  The log of the ionization parameter is 
labeled beside its corresponding curve. The cyan(black in hardcopy) highlights represent the 
regions for each ionization parameter where they are in the allowed parameter 
space.  {\bf(a)} Graph for   $n_e = 10^{11}$ cm$^{-3}$. {\bf(b)} Graph for   
$n_e = 10^{12}$ cm$^{-3}$.  [A colour version of this figure is available in the online version]
}
\label{Calcium2Graph}
\end{figure}

However, the UV-derived $N_H \simeq 3.11 \times 10^{22}$ cm$^{-2}$ is much less 
than the X-ray-derived $N_H \ge 1.8 \times 10^{24}$ cm$^{-2}$.  This can be 
explained by assuming a higher ionization parameter at the cloud face.  Using 
CLOUDY, we investigated the behavior of \caii\ with higher values for
the ionization parameter;  namely $\log U=0.5$,$1.0$,$1.5$, and $2.0$.  We 
also varied the density: separate runs for a density of $n_e=10^{11}$ cm
$^{-3}$ and $n_e=10^{12}$ cm$^{-3}$ were carried out for each of the above 
values of $\log U$. The results are shown in Figure \ref{Calcium2Graph}.

\begin{figure}
\subfigure{
\includegraphics[scale=0.35]{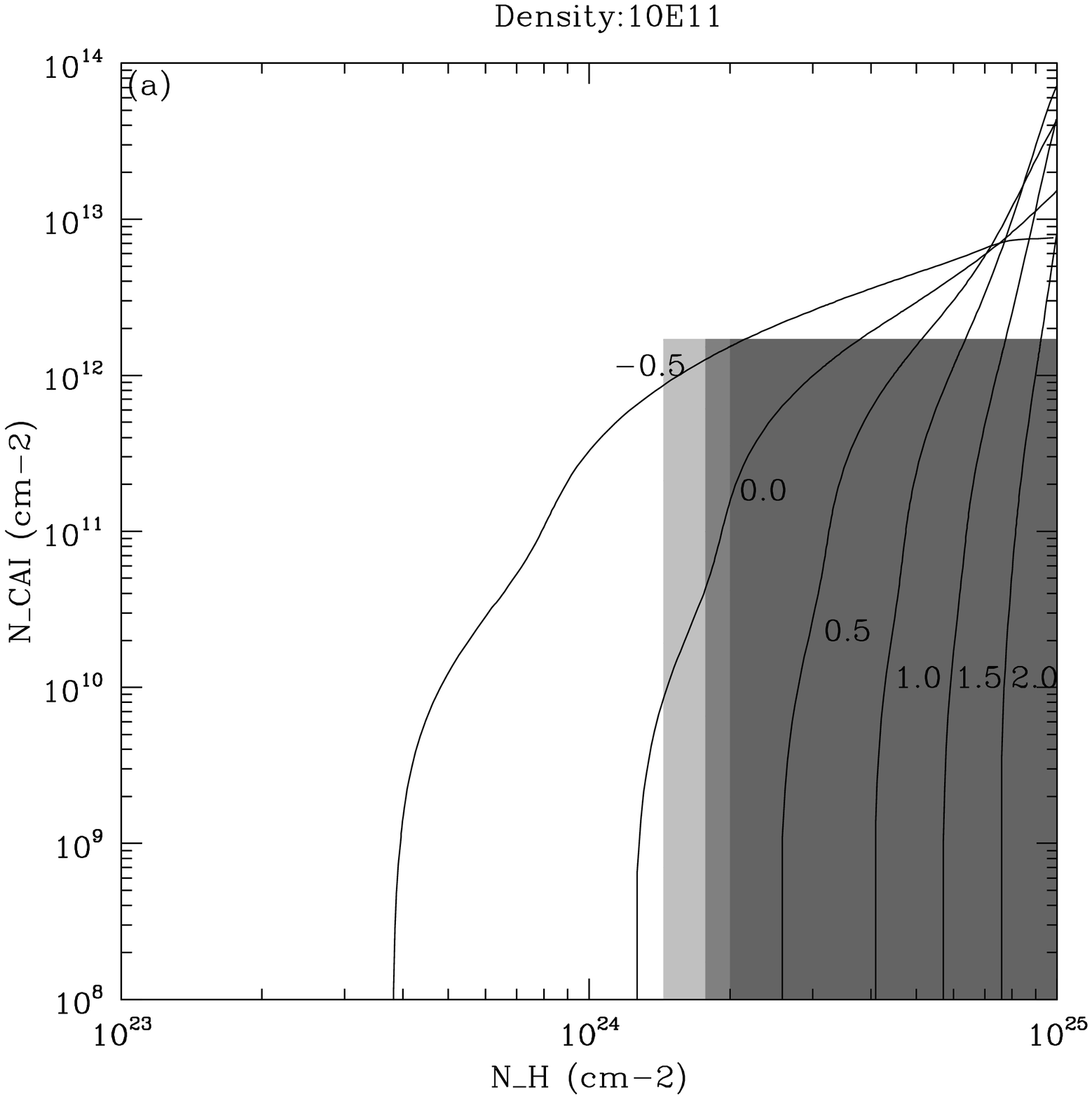}
}
\subfigure{
\includegraphics[scale=0.35]{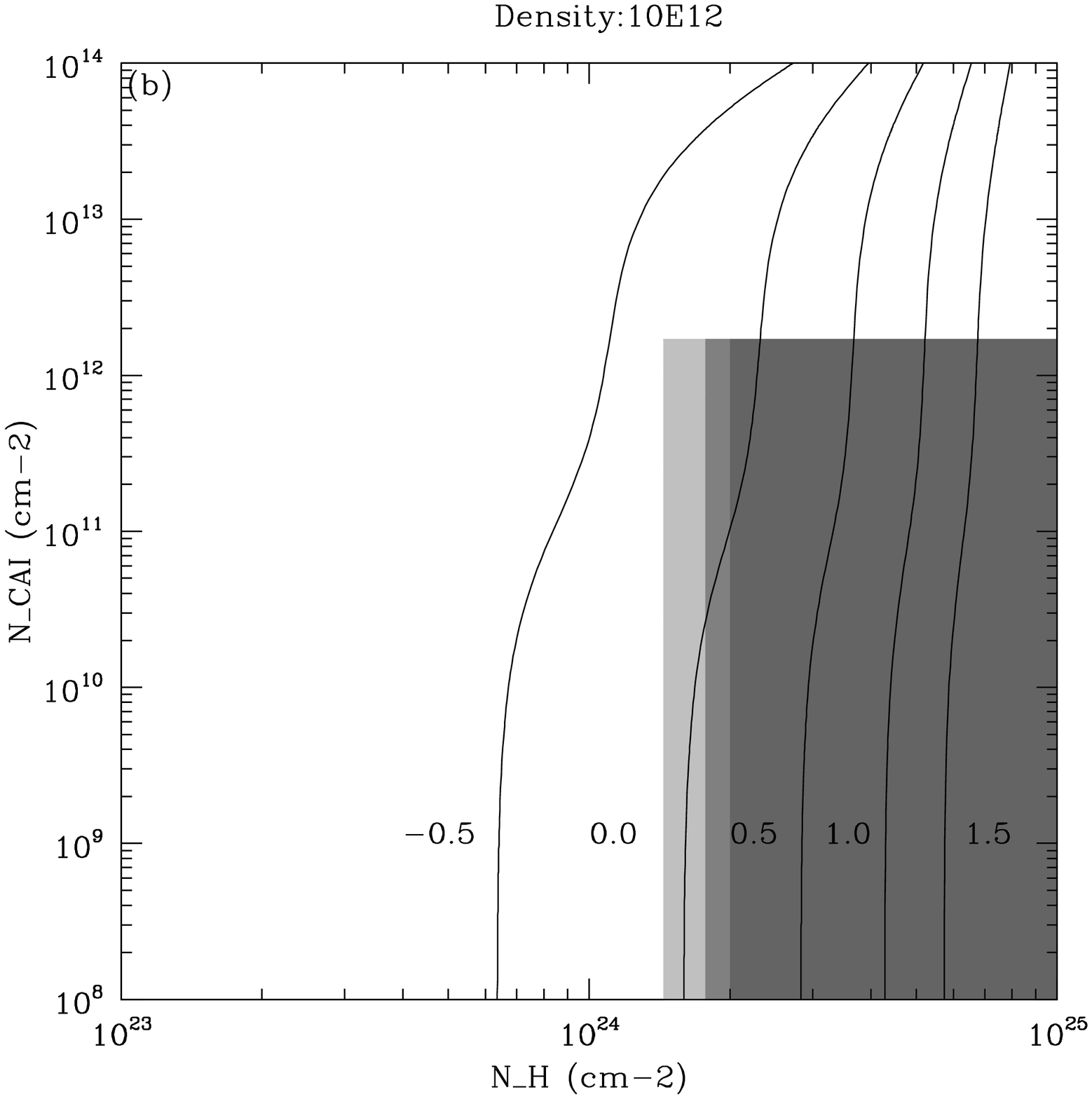}
}
\caption{These curves correspond directly to the curves seen in Figure 
\ref{Calcium2Graph}, as a check of how consistent the lower limit on the 
log of the ionization parameter is with other ions.  A hard upper limit 
(calculated via UV spectroscopy) is set for the \cai\ column density in \otbal.
The same
hydrogen column density is used for the vertically shaded regions. Again, the
curves are labeled with their corresponding ionization parameter.  {\bf(a)} 
Graph for $n_e = 10^{11}$ cm$^{-3}$. {\bf(b)} Graph for   $n_e = 10^{12}$ 
cm$^{-3}$.
 }
\label{Calcium1Graph}
\end{figure}

The shaded regions are defined by the UV-derived \caii\ column density 
(horizontal region) and the X-ray derived lower limit on the hydrogen column 
density (vertical regions).  The lower limit on the hydrogen column is shown at
1$\sigma$ (lightest grey), 2$\sigma$, and 3$\sigma$ (darkest grey).  The curves
represent the different $\log U$ of interest for this object.  For a specific
$\log U$ to match the observations, its curve must fall in the allowed region
for both the column densities plotted.  In Figure \ref{Calcium2Graph},
the allowed region for each $\log U$ is represented by a cyan 
highlight.  Thus, this object requires $\log U>0.0 $ for both $n_e=10^{11}$ 
cm$^{-3}$ and $n_e =10^{12}$ cm$^{-3}$ (Figures \ref{Calcium2Graph}a and 
\ref{Calcium2Graph}b, respectively).  This lower limit is consistent with 
the constraints posed by \cai\ (plotted in Figure \ref{Calcium1Graph}), and also the
constraints from absorption in \mgi, \mgii, and \feii.

However, as we shall see in \S\,\ref{discussotbal}, observations of \feii*
rule out the $\log U > 0$, $n_e \simeq 10^{11}$~cm$^{-3}$ scenario; gas
with such parameters may form part of the absorber in this object, but it
cannot form the entire absorber.

%---------------------------------END OTBAL----------------------------------%
%\clearpage

\section{Results: SDSS J2215$-$0045}	%1.4755
\label{febal}

%--tstop-tstart = 6615.86278
\febal\ was observed for 6615.9 seconds on June 21, 2003 (UT).  The measured 
background of the image is such that we expect only $0.102 \pm 0.004$ observed-frame 0.5-6.65 keV background photons 
in the detection apertures.

To find \fuv\ we fit a power law in $f_\nu$ to narrow normalization windows
centered at 2668~\AA\ and 3035~\AA\ rest-frame.  %{\bf We use a 10\%\ error on the UV flux for the following discussion.}
%We adopted the value of this power law at rest-frame 2500~\AA\ as \fuv. 
The result is \fuv\ 
$=(7.26 \pm 0.73) \times 10^{-27}$ erg s$^{-1}$ cm$^{-2}$ Hz$^{-1}$.
The true \fuv\ could be somewhat higher since we do not account for
possible \mgii\ absorption in the shorter-wavelength normalization region,
nor do we account for the inferred continuum reddening of \ebv$\simeq$0.06 
\citep{sdss199}.  Accounting for either of those effects would increase \fuv, 
which would increase the predicted \fx, which would in turn increase the 
absorption needed to bring the predicted \fx\ below the upper limit we measure.
Thus, this value of \fuv\ is conservative with respect to the amount of X-ray 
absorption we infer.

No photons are detected in the observed 0.5$-$6.65~keV bandpass, %(rest frame 1.24$-$16.46~keV)
so we have a conservative upper limit of 3.00 photons 
at 95\% confidence.\footnotemark[5]
This limit corresponds to a count rate of $\le 4.544 \times 10^{-4}$ photons 
s$^{-1}$.  The observed X-ray flux limit is therefore \fx $\le 1.776 \times 
10^{-33}$ erg s$^{-1} $ cm$^{-2}$ Hz$^{-1}$.  %(webPIMMS).  
From the observed 
X-ray flux upper limit we calculate the observed \aox\ upper limit, using 
equation (\ref{alphaoxobs}), and find $\alox \le -2.45$ at 95\%\ confidence.

\footnotetext[5]{One photon with energy 12.7~keV is detected
within 2\farcs5 of the position of \febal, but photons with such high energies
are much more likely to be background photons, as discussed in \S \ref{Xray}.}

The expected X-ray flux is $\log$(\fx) $=-30.53^{+0.78}_{-0.77}$ erg s$^{-1}$ 
cm$^{-2}$ Hz$^{-1}$, which includes the uncertainty
in the power law slope in the UV band.  
Given that flux, $2.70^{+0.77}_{-0.80}$ counts in the log should be observed
over 6615.9s, which is higher than the observed 95\% confidence limit of 
$0.5$ in the log.  We also expect $\aloxUV=-1.74 \pm 0.35$. 
The deviation of the observed \aox\ from the predicted value 
is $\Delta \alox \le -0.71$.  Compared to \otbal, we probe higher-energy
photons in \febal\ due to its higher redshift, and so to produce the same 
observed $\Delta \alox \le -0.71$ requires a higher absorbing column:  $N_H \geq 3.4 \times10^{24}$ cm$^{-2}$ at the quasar redshift, assuming solar metallicity.
Again, this is a hard lower limit assuming a neutral absorber.

\citet{sdss5BAL} also evaluated values for this object.  They find $\alox < -2.71$ and $\Delta \alox\le -0.96$.  Again, due to slightly different formulae, the values derived in \citet{sdss5BAL} are slightly different than those calculated here.  In this case, the $\Delta \alox$ value is $< 1.5 \sigma$ different.

%--------------------------END Calculations-----------------------------------%
The presence of doubly ionized iron (\feiii) absorption without any significant singly ionized iron (\feii) absorption is a rare occurrence.  \citet{dek02b} showed that this iron absorption behaviour could be produced with a density of $n_e=10^{11}$ cm$^{-3}$, an ionization parameter $\log U=-1.5$, and a very narrow range of total hydrogen column density around $\log N_H \sim 22.4$.  We investigated this result with 
%the addition of the UV derived values for $N_{\rm FeIII}$ and $N_{\rm FeII}$ and the X-ray derived value for $N_H$ as constraints in 
CLOUDY simulations using the same parameters as for \otbal.  The results for \feiii\ are located in Figure \ref{Iron3Graph} and the results for \feii\ in Figure \ref{Iron2Graph}.

In Figure \ref{Iron3Graph},
the shaded region is shaped by the lower limits of both the X-ray 
derived hydrogen column (vertical lower limits) and the UV-derived 
\feiii\ column density (horizontal lower limits; details given in 
\S\,\ref{discussfebal}).  The curves must fall inside 
this region to correctly describe the observations.  
The curves must also satisfy the upper limit on the \feii\ 
column density (also discussed in \S\,\ref{discussfebal})
shown in Figure \ref{Iron2Graph}.  
%To correctly describe a system with \feiii\ absorption and 
%little \feii\ absorption.  
%To take this into account, the ionization 
%parameters are also plotted for the \feii\ column in Figure \ref{Iron2Graph}.  
The shaded region, in this case, is defined by an upper limit \feii\ column.
%To match the observations, the curve must fall in both the shaded regions of
%Figure \ref{Iron3Graph} and \ref{Iron2Graph}. 

The cyan segments in Figure \ref{Iron3Graph} represent the interval
over which each curves match the observations by falling within the shaded
regions of both Figures \ref{Iron3Graph} and \ref{Iron2Graph}.
We can place a lower limit of $\log U>1.0$ on the ionization
parameter for both the $n_e=10^{11}$ cm$^{-3}$ and $n_e
=10^{12}$ cm$^{-3}$ cases (Figures \ref{Iron2Graph}a and 
\ref{Iron2Graph}b, respectively).  This lower limit is consistent with 
constraints imposed by \mgii, \alii, and \aliii.

The physical basis behind these constraints is that
\feiii\ absorption without significant \feii\ absorption is seen
when the absorber has just barely enough column density to form a hydrogen
ionization front, because \feiii\ is only abundant just before such a front and
\feii\ is abundant just after one.  Furthermore, to create an \feiii\ column
matching the observations in the narrow region just before the front
requires a high density, because such high densities
force \feiv\ to recombine to \feiii\ \citep{bp98}.

%--------------------------END OF \FEBAL--------------------------------------%
%\clearpage

\section{Discussion}
\label{discuss}

We have shown that both \otbal\ and \febal\ are undetected in snapshot X-ray
observations.  Despite the short exposure times of the X-ray observations,
non-detections of these optically bright quasars require very large
absorbing columns and small levels of unabsorbed X-ray scattering
($<$1.3\% at 95\% confidence).\footnotemark[6]  
Note that our column density lower limits assume neutral absorption.  
As shown by \citet{SBCG10}, when 
an ionized absorber model is used, BAL quasar X-ray absorbing
columns can be 1-2 orders of magnitude larger than in the neutral case.
Thus, the column densities we quote here are hard lower limits, and 
the true column densities are almost certainly larger.
We now discuss the implications of these results in more detail.

\footnotetext[6]{These X-ray scattering constraints 
are broadly consistent with optical polarization measurements for both objects
(Hall, Smith, et al., unpublished): \otbal\ has $R$-band $P=1.58\pm0.10$\% and
\febal\ has a white-light $P=0.40\pm0.06$\%.}

\subsection{SDSS J0300+0048}
\label{discussotbal}

In \otbal, at outflow velocities of $2000<\msv<4000$~\kms\ we see 
\mgii, \mgi, \caii\ and \feii\ absorption.
At $4000<\msv\lesssim 10850$~\kms, 
we see \mgii, \feii\ and excited-state \feii\ (\feii*) absorption from 
gas which must be located closer to the quasar than the 
\caii-absorbing region.  (If it were located farther from the quasar than the 
\caii-absorbing region, it would be shielded by that gas and would itself show 
\caii.)  

\cite{sb2} and \cite{pbhmhd} suggested that the \otbal\ outflow 
could be produced by gas in an accretion disk wind if
we are looking across the wind rather than down the wind.  
In this model, high-ionization gas is launched
from closer to the quasar than low-ionization gas is, with \caii\ and
\mgi\ being launched from outside 
a hydrogen ionization front.\footnotemark[7]
%	For at least the low-ionization gas (seen in \mgii\ and \caii),
%	our line of sight intersects the outflow
%	only after it has been accelerated to $\gtrsim 2000$~\kms.  

\footnotetext[7]{The radial dependence of the density in the outflow at launch
is unclear.
As mentioned in \cite{sb2}, in region (a) of a \cite{ss73} disk the density
increases with radius.  A disk wind launched from that region would have a
higher density in the \caii\ region than in the \feii* region.  
A low temperature ($T \lesssim 1100$~K)
would be required in the \caii\ region to avoid exciting \feii* there.
On the other hand, the low-ionization absorption in this
object covers the \mgii\ (and \feii) broad emission line region,  %(BELR).
located at radii where the \cite{ss73} disk model 
predicts a radially decreasing density.
If the disk wind is launched at these larger radii and its density is
decreasing with distance, the density in the \caii-absorbing region 
%(located farthest from the quasar) 
should be the lowest in the outflow.
A density of $n_e\lesssim 10^3$~cm$^{-3}$ would be required to avoid
populating the lower levels of \feii* transitions which are not observed.
}

We can constrain the density in the \feii*-absorbing region by studying the 
critical densities of excited levels from which we either see or do not see
\feii* absorption.  We see absorption from the \feii\ UV78 multiplet, whose 
lower level has
a critical density of $\simeq 10^{5.5}$~cm$^{-3}$.  We do not see absorption 
from \feii\ UV195/UV196 multiplets, with relevant critical densities of
$\simeq 10^{7.5}$~cm$^{-3}$.  Thus, we can constrain the density in the 
\feii*-absorbing region to be $\simeq 10^{6.0\pm 0.5}$~cm$^{-3}$.  
%We do not see absorption from \feii\ Opt105, which has a critical density of 
%$\simeq 10^{11.5}$~cm$^{-3}$.  

We now examine the ionization behaviour of a slab of gas at this density,
to see if a single-slab model can explain both the X-ray and UV absorption.
In Figure \ref{lowdensitycal} we present two CLOUDY runs for 
$n_e = 10^6$~cm$^{-3}$, with $\log U = 1.0$ and $\log U = -0.5$.
%The latter CLOUDY run halted when it reached a temperature $T_e=4000$\,K, so the curve for $\log U = -0.5$ shown was forced to have a stopping temperature of $T_e=1000$~K.
%
For $\log U = 1.0$, the observed \feii\ column (Figure \ref{lowdensitycal}a)
can be reached with reasonable hydrogen column densities, but the \caii\ column
(Figure \ref{lowdensitycal}b)
can only be achieved with $N_H \sim 10^{25}$~cm$^{-2}$.  
When the hydrogen column approaches $\sim 10^{25}$~cm$^{-2}$ 
between us and the quasar UV emission region,
the electron scattering optical depth rises above $\tau_{es} \sim 2$,
increasing the already high intrinsic optical and UV
luminosity of this object.
(Of course, the X-ray absorption may cover only the X-ray continuum-emitting
region, and not the larger UV continuum-emitting region, but at the moment we
are only considering whether a single uniform absorber can explain the data.)
For $\log U = -0.5$, the observed \caii\ column (Figure \ref{lowdensitycal}b) is 
reached at $\sim 3\times 10^{24}$~cm$^{-2}$, at which point the predicted
\feii\ column is well above the observationally inferred lower limit 
and the temperature within a factor of two 
of the observationally inferred upper limit.  

Thus, in light of the X-ray data, we find that we can nearly match 
the properties of both the X-ray and UV absorbers in 
\otbal\ with one slab of gas of constant density $n_e = 10^6$~cm$^{-3}$
with $\log U = -0.5$ at its ionized face, implying an absorber 
at a distance of $\sim$60 pc from the black hole
with a thickness of $\sim 3 \times 10^{18}$~cm or $\sim$1~pc 
%If the gas density increases with
%distance, so that it is higher in the \caii\ region,
%the temperature there should be even lower, confirming
%the plausibility of the disk wind scenario first outlined in \cite{sb2}.

%The thickness of the outflow may favor a disk wind model where the wind is
%launched from a wide range of radii around the black hole.  For a 10$^9$
%$M_\odot$ black hole typical of SDSS quasars at $0.7<z<2.2$, the circular
%velocity 1 pc from the black hole is of order 2000 km~s$^{-1}$, which is
%the minimum velocity observed for the outflow in this object.  Launching
%the observed outflow over a radial extent of $\sim$1 pc is not a problem,
%but keeping such an outflow collimated over $\sim$60 pc may be.
%
%If the density instead decreases with distance, the density at the
%hydrogen ionization front surrounding the quasar
%(outside which \caii\ is seen) would have to be $\sim 10^3$~cm$^{-3}$
%to ensure that no \feii* absorption occurs outside it.  Detailed models
%for this latter scenario are beyond the scope of this work.

If the absorbing gas in \otbal\ can be approximated by gas with a uniform
density of $n_e = 10^6$~cm$^{-3}$, it is difficult to reconcile the resulting
$\sim$60 pc distance of the gas from the black hole with the original
disk wind scenario of \cite{sb2}.  Such a distance is more consistent with
the alternative scenario to a disk wind discussed in \cite{sb2}; namely,
that the outflow sweeps up gas at large distances
from its origin and decelerates in the process.
In this swept-up-gas scenario, the \feii*-absorbing gas would be gas
which has been compressed to $n_e\gtrsim 10^3$~cm$^{-3}$
and accelerated to $4000<\msv<10850$~\kms\ by a
high-ionization, low-density wind (seen at velocities up to at least
$\msv=10850$~\kms\ in this object).
The \caii-absorbing gas would be gas which has been overtaken by the wind more
recently and which has therefore been compressed and accelerated to a lesser
degree than the \feii*-absorbing gas.
However, it remains to be seen 
(through detailed outflow modeling beyond the scope of this investigation)
whether gas can be accelerated to $2000<\msv<4000$~\kms\ without being
compressed to a density $n_e>10^3$~cm$^{-3}$
or heated to a temperature $T>1100$~K.
Meanwhile, as suggested by \cite{sb2}, this swept-up-gas scenario can be
tested by looking for a long-term increase of the \caii\ BAL outflow velocity.
(No increase was seen over a rest-frame time span of up to 205 days in multiple
SDSS spectra plus our VLT spectrum of this object.)

%\clearpage

\subsection{SDSS J2215$-$0045 }
\label{discussfebal}

The conservative upper limit of $\log N_{\rm FeII}<14.7$ (Figure 
\ref{Iron2Graph}) for this object was estimated using the plausible unabsorbed
continuum from \citealt{sdss199} (hereafter HH03), shown in Figure 1b.  
The details of the estimation of the lower limits for the \feiii\ column density 
shown in Figure \ref{Iron3Graph} are given in the Appendix.

We have found in \S \ref{febal} that if a single absorber
is responsible for both the observed UV and X-ray absorption in \febal, 
it must have $N_H>3.4 \times 10^{24}$ cm$^{-2}$, 
$\log n_e\gtrsim 9.5$ cm$^{-3}$ and $\log U\geq 1$,
with full coverage of the X-ray source 
and at least 50\% coverage of the UV continuum source.
We now consider whether an absorption system with those properties is in fact
plausible.

If we take $N_H=3\times 10^{24}$ cm$^{-2}$ and $\log n_e=9.5$ cm$^{-3}$,
then the absorber is only 10$^{15}$ cm thick.  
A larger $N_H$ would make for a thicker absorber but would also increase the
electron scattering optical depth above $\tau_{es}\simeq 2$, 
increasing the already high intrinsic luminosity of this quasar.
%(Even $\tau_{es}\simeq 2$ likely requires a special viewing geometry.)
The X-ray continuum region is $\sim$10$^{15}$ cm in radius \citep{CK09}, while the UV continuum
region is $\sim$10$^{16}$ cm in radius \citep{KD07}.
To cover both regions, the absorber would have to be at least 20 times
larger in the transverse direction than it is in the line-of-sight direction.
Moreover, the line-of-sight velocity width of the absorber is 12,000 km 
s$^{-1}$.  If that is a turbulent $\Delta \msv$, the absorber would double its 
thickness and halve its density on a timescale equal to the crossing time of 
$\sim$10 days, but observations of \febal\ over rest-frame timescales much
longer than that reveal no changes in its absorption troughs (HH03).
If that $\Delta \msv$ is a coherent velocity spread along our 
line of sight, the acceleration required to produce it in 10$^{15}$ cm is 
$\sim$100 m\,s$^{-2}$.  For comparison, in the disk wind model of \cite{mcgv}
the radiative acceleration is $\lesssim$100 cm\,s$^{-2}$.

It is hard to see how to bring the above extreme parameters
into the realm of the plausible.  For example,
a supersolar Fe abundance would reduce the $N_H$ required to match the
observed $N_{\rm FeIII}$, and Ly$\alpha$ pumping of \feiii\ UV34 \citep{jea00}
would reduce the amount of Fe needed to explain a given $N_{\rm FeIII}$.
Those effects could eliminate the $\tau_{es}$ problem, but would shrink the
line-of-sight width of the absorber and make the dynamical problems worse.

Thus, a single, uniform absorber with the properties required to explain the
UV and X-ray absorption in this system is extremely unlikely to exist.
Instead, it is likely that the X-ray absorber is Compton-thick
($\gtrsim 3\times 10^{24}$ cm$^{-2}$) but compact (radius $\sim 10^{15}$ cm),
so that at most a small part of the UV continuum region is obscured by it.
The UV absorber must be larger (radius $\sim 10^{16}$ cm) and have a 
column density just under that required to form a hydrogen ionization front
($N_H \lesssim U\times 10^{23}$ cm$^{-2}$) so that it contains very little
\feii.  We can constrain $-1\lesssim \log U\lesssim 1$ for the UV absorber.
The upper limit comes from setting the electron scattering optical
depth to the UV continuum source to be at most unity.  The lower limit comes
via our measured $\log N_{\rm FeIII}>16.62$ and $\log N_{\rm FeII}<14.7$; 
given those values, Table 2 of \citet{dek02b} and our Figures \ref{Iron2Graph}
and \ref{Iron3Graph} show that $\log n_e\gtrsim 9.5$ cm$^{-3}$ and 
$\log U\gtrsim -1$.\footnotemark[8]

\footnotetext[8]{We 
assume that the UV absorber is fully shielded by a dust-free X-ray
absorber.  In that case, the X-ray
absorber will absorb a relatively small fraction of hydrogen-ionizing photons,
and the structure of the absorber's Stromgren sphere will be much less
affected than the structure of its partially ionized zone.
We can therefore use our constant-density CLOUDY simulations to infer
the conditions of the \feiii\ clumps, even though the higher-ionization
regions in the outflow must have a lower density.}

We can reconcile the high density required for the \feiii\ column
with the large velocity and velocity spread of \feiii\ if we assume the
\feiii\ absorption arises in dense clumps embedded in a lower density,
higher ionization wind seen in \civ\ and \SIiv, along the lines suggested
by \citet{vwk93}.  The clump widths along the line of sight
would be $d<N_H/n_e$, or $d<10^{13.5\pm 1.0}$ cm.  
This maximum size is larger than the typical size of 10$^{12}$ cm
predicted for putative BLR clouds \citep{kea97}.
To cover half the $r \sim 10^{16}$ cm UV source, 
a large number of such clumps ($\sim 10^{5\pm 2}$ if the clumps are roughly 
spherical) are required.

In this object,
\feiii\ absorption is seen at $6000<\msv_{\rm FeIII}<18000$~\kms\ along our
line of sight, whereas \civ\ and \SIiv\ absorption is seen at 
$0<\msv<25000\pm3000$~\kms\ (Figure 1 of HH03).
%the uncertain $\msv_{max}$ is due to the uncertain intrinsic continuum shape).
%
Given that the troughs %\feiii\ trough 
in this object have %has a 
higher outflow velocities %velocity 
and velocity widths %width 
than the average BAL trough, it is reasonable to assume that much
of the acceleration of the gas occurs along our line of sight; in
other words, the outflow streamlines are largely parallel to our line of sight.
If we imagine an accelerating, relatively low density wind (seen in \civ\ and
\SIiv) colliding with dense \feiii\ clumps, the wind will form a shock around
the clumps and can in principle accelerate, ablate, compress and bypass 
the clumps. 
%with the extent of each effect being dependent on the filling factors,
%densities, pressures and relative velocities of the wind and the clumps.
The acceleration may help explain the \feiii\ trough velocity, along with
radiative and possibly magnetohydrodynamic acceleration;
the ablation helps explains the \feiii\ trough width;
the compression helps explain the inferred high densities;
and the bypassing explains the \civ\ and \SIiv\ trough extending to both
smaller and larger outflow velocities than the \feiii\ trough \citep{vwk93}.

In summary, our picture of this quasar is that the X-ray-emitting region is 
completely covered by an X-ray absorber and that the more distant UV-emitting 
region is nearly or completely covered by a UV-absorbing wind seen in \civ\ and
\SIiv\ absorption.  This wind is accelerated (at least in part along our 
line of sight) from 0~\kms\ to $25000\pm3000$~\kms.  At some distance or range 
of distances from the quasar, the wind includes dense, \feiii-emitting clumps 
(with overall covering factor $\sim$50\%\ and low volume filling factor) which have been accelerated by the 
wind to velocities of $6000<\msv_{\rm FeIII}<18000$~\kms.  A paraboloidal shock 
surrounds each clump, where the wind gas encounters the slower-moving clump 
gas.  The shock-heated gas is no longer visible in \civ\ or \SIiv, but 
between a clump and its shock lies a region where gas ablated from the clump 
will absorb in those ions, at velocities $\msv\simeq \msv_{\rm FeIII}$.  The 
wind downstream from the clumps --- consisting of shocked gas and, if the volume 
filling factor of the clumps is small, unshocked gas as well --- is not seen to
recombine to observable ionization stages when shadowed by a clump, so its
recombination timescale must be longer than the time it spends in a shadow.

In this picture, the \feiii-emitting clumps are being ablated and so the
\feiii\ trough depth will decrease with time unless the wind encounters new
clumps.  Such clumps should first appear at low velocity; from their absence,
we conclude that the wind is not currently encountering new clumps.
Spectroscopic monitoring of the quasar to look for weakening \feiii\ troughs
(as well as any changes in \feiii\ trough velocity or velocity width)
would be worthwhile.  
%However, such a search could be complicated by trough
%variability caused by wind and cloud motions transverse to the line of sight.
%No significant variability has been seen to date in the troughs in this object
%(HH03). %\citep{sdss199}.

Lastly, we note the possibility that
this quasar may have developed from a typical FeLoBAL.  In such an object,
\feii\ is seen at low velocities and \civ\ at both high and low velocities, 
consistent with a high-ionization wind which has encountered clouds optically
thick in the Lyman limit and has only begun to accelerate them.  As time
progresses, the clouds will be accelerated and ablated.  Eventually, they may
be optically thin in the Lyman limit (and thus seen in \feiii\ and not \feii)
and will be found at higher velocities.

%In order of distance from the BH:
%X-ray src
%X-ray abs w/lo v
%UV src
%(invis UV abs?, v<6000 km/s)
%UV abs w/lo density & all v
%UV abs w/lo density + \feiii\ clumps with high density, >=50% covering 
%UV abs w/lo density & v~<v_feiii (slowed) + ~>v_feiii (bypassed)
%If the UV absorber is sufficiently close to the UV continuum region,
%parts of it might not be shielded from the X-ray source by the X-ray absorber,
%but that's unlikely.

%Assuming $N_{\rm FeIII}>10^{16.62}$ and a trough width of 12,000~\kms,
%the average $\log N_{\rm FeIII}(v)=13.5$ cm$^{-2}$ per 10~\kms\ velocity
%bin, where 10~\kms\ is a typical thermal velocity for $\sim 10^4$~K gas.
%Given a solar iron abundance of -4.5 in the log, this $N_{\rm FeIII}(v)$
%translates to an average $\log N_H(v)=10^{18}$ per 10~\kms\ velocity bin.

%-------------------------END OF DISCUSSION-----------------------------------%
%\clearpage

\section{Conclusion}
\label{conclusion}

\otbal\ and \febal\ have some of the highest known column densities in their BAL outflows.

\otbal\ specifically has a very high \mgii\ absorbing column, which creates the overlapping absorption below rest-frame 2800\AA.  Exploiting the correlation between X-ray and UV luminosities, we find $\Delta \alox < -0.72$, indicating strong X-ray absorption.  The X-ray absorbing gas must have a column density of $N_H \geq 1.8 \times 10^{24}$ cm$^{-2}$ (assuming solar metallicity).  
Absorption from \feii\ UV78 but not \feii\ UV195/UV196 in the UV spectra indicates a density in this absorbing region of $n_e \simeq 10^{6}$ cm$^{-3}$.  We have provided two CLOUDY runs at this density with ionization parameters $\log U = 1.0$ and $\log U = -0.5$.  Most properties of the absorber in \otbal\ can be matched with a slab of gas of constant density $n_e = 10^6$~cm$^{-3}$ with $\log U = -0.5$ at its ionized face (implying a distance of $\sim$60~pc from the black hole) and a thickness of $3 \sim 10^{18}$~cm ($\sim$1~pc).

For \febal, we showed that a single absorbing slab of density $\log n_e \ge 9.5$ cm$^{-3}$ and $\log U \ge 1$ that fully covers the X-ray source and covers at least 50\%\ of the UV continuum could explain the total hydrogen column and \feii+\feiii\ column densities.
These parameters are shown to have little physical plausibility, as they require an abnormally high acceleration to produce the observed line-of-sight velocity spread seen in the UV absorption troughs (the velocity spread would otherwise disperse the slab and reduce its density on extremely short timescales).
One possible scenario to explain the absorption would be dense \feiii-emitting clumps inside a lower density UV-absorbing wind (seen in \civ\ and \SIiv\ absorption).
The UV wind is accelerated from 0 km s$^{-1}$ to 25000$\pm3000$ km s$^{-1}$, and impacts the the denser \feiii\ clumps.
The clumps are accelerated to the observed velocity dispersion while being ablated by the UV wind.
Such a model could be tested by looking for evaporation of the dense \feiii\ absorbing clumps manifesting itself as time-varying absorption features in the UV spectra.
Explaining the rarity of \feiii-dominant BAL quasars is easier: only a small parameter space for strong \feiii\ absorption with little \feii\ absorption present is found by both \citet{dek02b} and our own CLOUDY runs.

%------------------------ END OF CONCLUSION-----------------------------------%

We thank M. Bautista for discussions regarding \feiii.
PBH acknowledges the hospitality of the University of Wyoming Department of
Physics and Astronomy and the Aspen Center for Physics, where parts of this
work were completed.
PBH acknowledges financial support from a Fundaci\'{o}n Andes grant and from
NSERC; PBH and MSB acknowledge financial support from Chandra (proposal
04700849).
This research made use of the Atomic Line List v2.04 at 
\url{http://www.pa.uky.edu/~peter/atomic/}.

% REFERENCES -----------------------------------------------------------------%
\footnotesize
\bibliographystyle{elsarticle-harv}
%\bibliographystyle{elsarticle-num-names}
%\bibliography{/Users/phall/OLDWORK/AASTEX/pathall}
%\bibliography{/home/phall/OLDWORK/AASTEX/pathall}
%\bibliography{/home/jar/project1/pathall}

%%TABLES ---------------------------------------------------------------------%

%%FIGURES / FIGURE CAPTIONS --------------------------------------------------%
%------------------------------OTBAL-----------------------------------%
\clearpage

%\begin{figure}
%\centering
%\includegraphics[scale=0.9]{otbalfig.eps}
%\caption{Partial SDSS spectrum of SDSS J0300+0048.  The vertical axis
%gives $F_\lambda$ in units of 10$^{-17}$ erg s$^{-1}$ cm$^{-2}$ \AA$^{-1}$.
%The bottom horizontal axis gives the observed wavelengths in \AA;
%while the top horizontal axis gives the rest-frame wavelengths in \AA.
%Emission lines are identified in the center of the plot,
%narrow associated lines at the systemic redshift along the bottom,
%and BAL troughs along the top.
%Below 2800\,\AA\ in the rest frame, \feii\ and other troughs overlap
%to absorb nearly all the flux from the quasar; hence, the quasar is
%termed an ``overlapping-trough'' BAL quasar \citep{sdss123}.
%}
%\label{OTBAL}
%\end{figure}

%------------------------------FEBAL-----------------------------------%
%\begin{figure}
%\centering
%\includegraphics[scale=0.75]{febalfig.ps}
%\caption{Spectra of the \feiii-dominant BAL quasar SDSS J2215$-$0045 
%in the rest frame, from \citet{sdss199}.  
%Green shows the SDSS spectrum, black a binned UVES spectrum,
%and red an estimate of the intrinsic, unabsorbed spectrum.
%The UV34 and UV48 multiplets of \feiii\ are very strong in absorption,
%while only upper limits can be put on absorption from any \feii\ multiplet.
%}
%\label{FEBAL}
%\end{figure}

%-----------------------------CalciumII Graph---------------------------------%
%------------------------------CalciumI Graph---------------------------------%
%\clearpage

\clearpage
%------------------------------IronIII Graph----------------------------------%
\begin{figure}
\subfigure{
\includegraphics[scale=0.33]{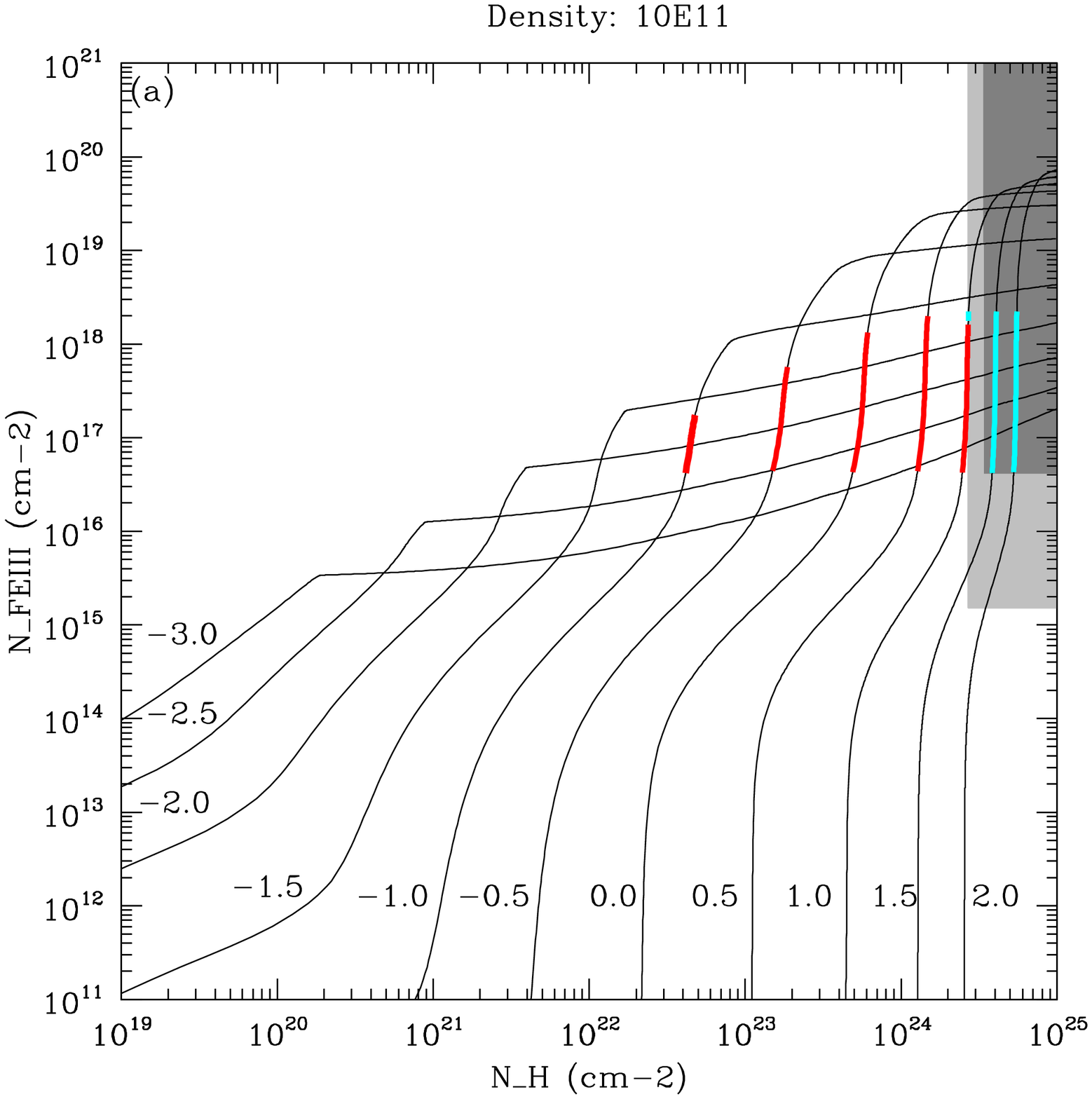}
}
\subfigure{
\includegraphics[scale=0.33]{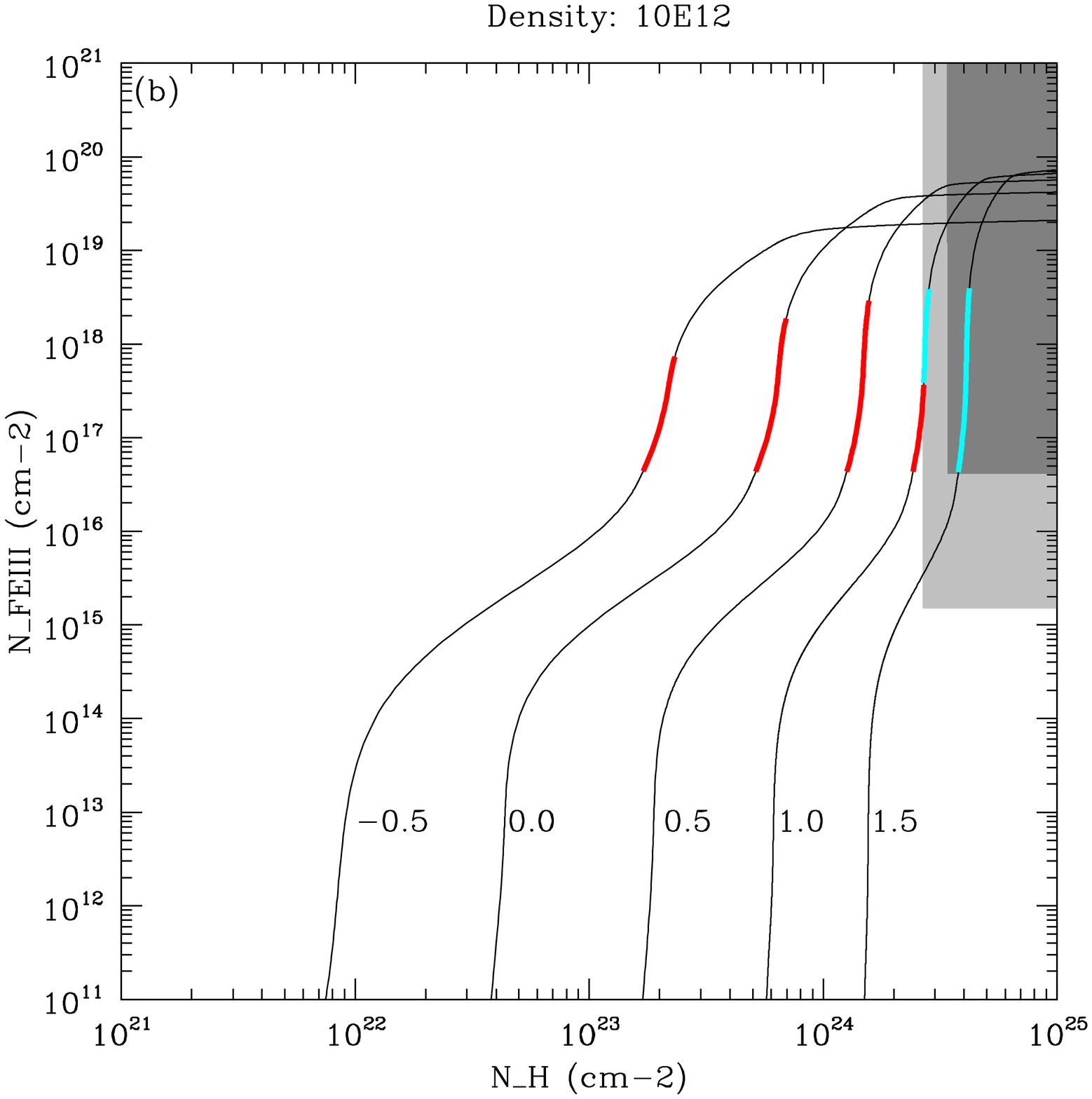}
}
\caption{Curves represent the behaviour of the \feiii\ column density for 
varying ionization parameter and density.   The lower limit on the hydrogen 
column in \febal\ 
is shown at 1$\sigma$ (lightest grey) and 2$\sigma$ (darker grey).  An 
uncorrected (lightest grey) and corrected (darker grey) lower limit on the 
\feiii\ column is presented (see \S \ref{discussfebal}).  The log of the 
ionization parameter is labeled beside its corresponding curve. The cyan (black in hardcopy) 
highlights denote the regions for which the curves satisfy both the lower limit
\feiii\ column density and the lower limit hydrogen column density as well as
the upper limit \feii\ column density (see Figure \ref{Iron2Graph}).  Red(light grey in hardcopy)
highlights only satisfy the constraints posed by the \feiii\ and \feii\ column
densities, disregarding our lower limit on the hydrogen column.  {\bf(a)} 
Graph for $n_e = 10^{11}$ cm$^{-3}$. {\bf(b)} Graph for   $n_e = 10^{12}$ 
cm$^{-3}$. [A colour version of this figure is available in the online version]
}
\label{Iron3Graph}
\end{figure}

%------------------------------IronII Graph-----------------------------------%
\begin{figure}
\subfigure{
\includegraphics[scale=0.33]{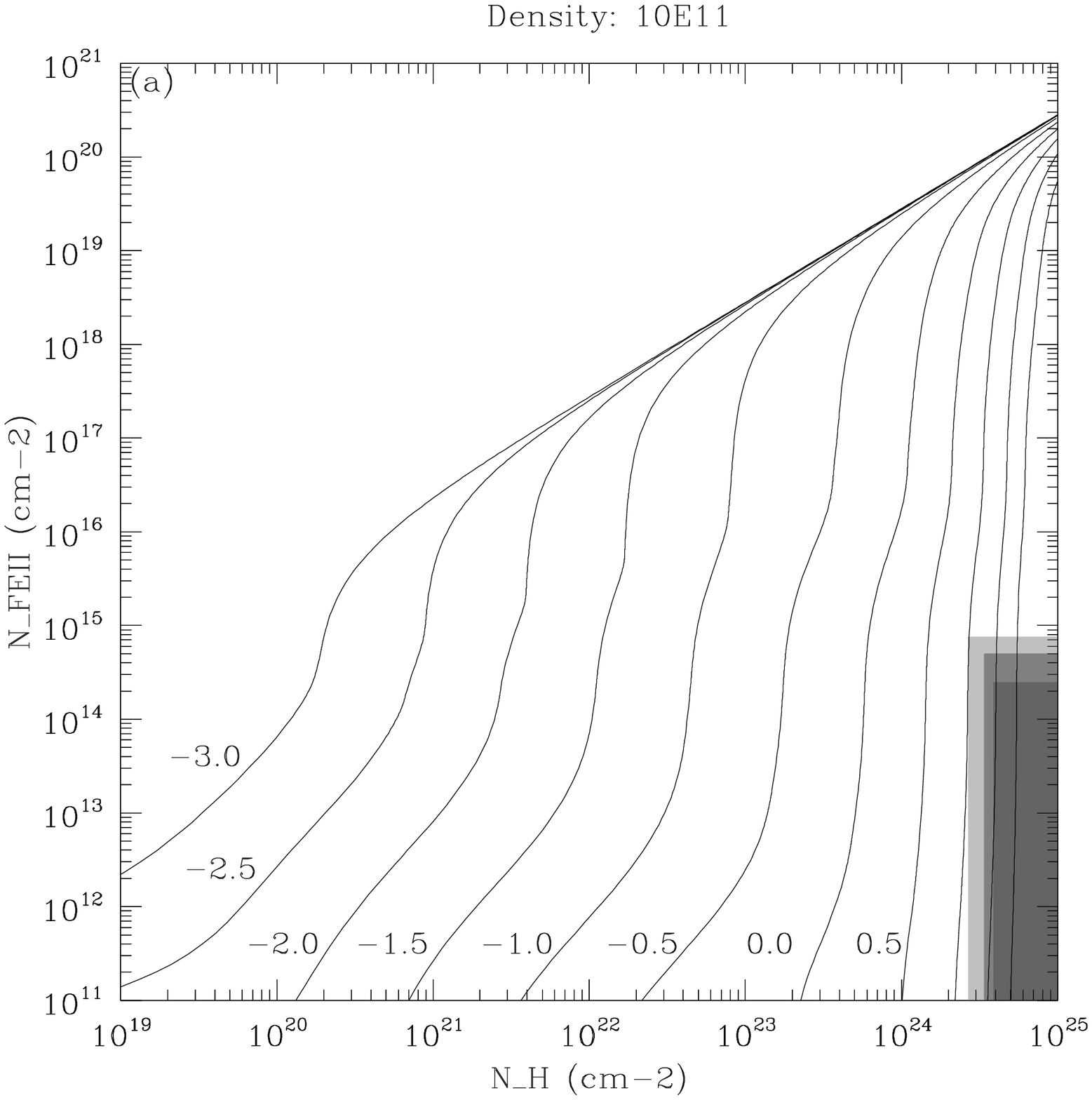}
}
\subfigure{
\includegraphics[scale=0.33]{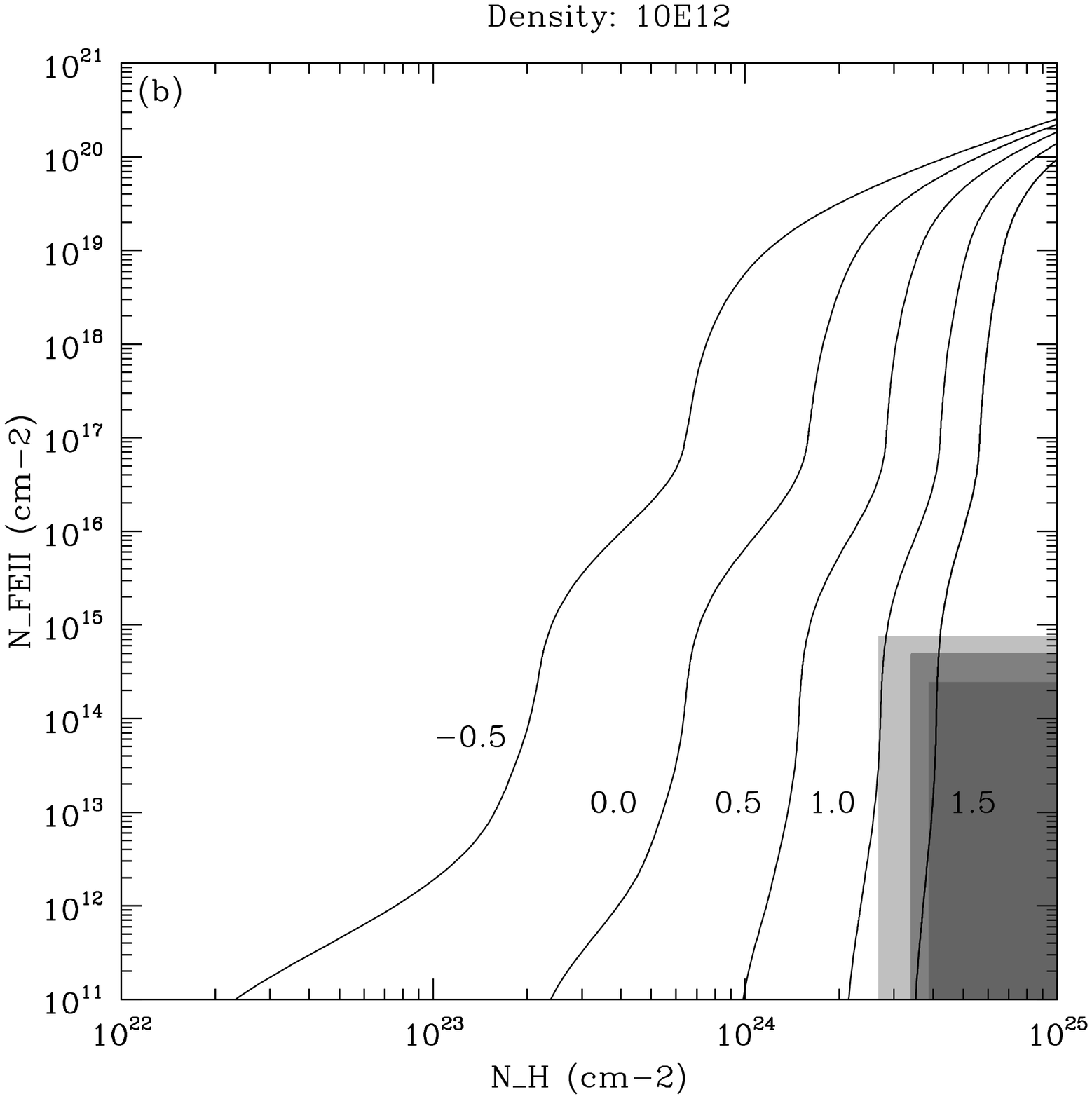}
}
\caption{The relationship of \feiii\ and \feii\ absorbing gas in this object is vital. Here we plot the same curves for each ionization parameter as seen in Figure \ref{Iron3Graph}, but in comparison to the upper limit on \feii\ absorbing column (see \S\ \ref{discussfebal}).  The lower limit on the hydrogen column density is also the same as seen in Figure \ref{Iron3Graph}.  These plots show that the ionization parameters supported by the lower limit on the \feiii\ column are also supported by the upper limit on the \feii\ column density.  {\bf(a)} Graph for   $n_e = 10^{11}$ cm$^{-3}$. {\bf(b)} Graph for  $n_e = 10^{12}$ cm$^{-3}$. }
\label{Iron2Graph}
\end{figure}

%-----------------------------CAII/FEII for log nH = 6-----------------------%
\begin{figure}
\subfigure{
\includegraphics[scale=0.35]{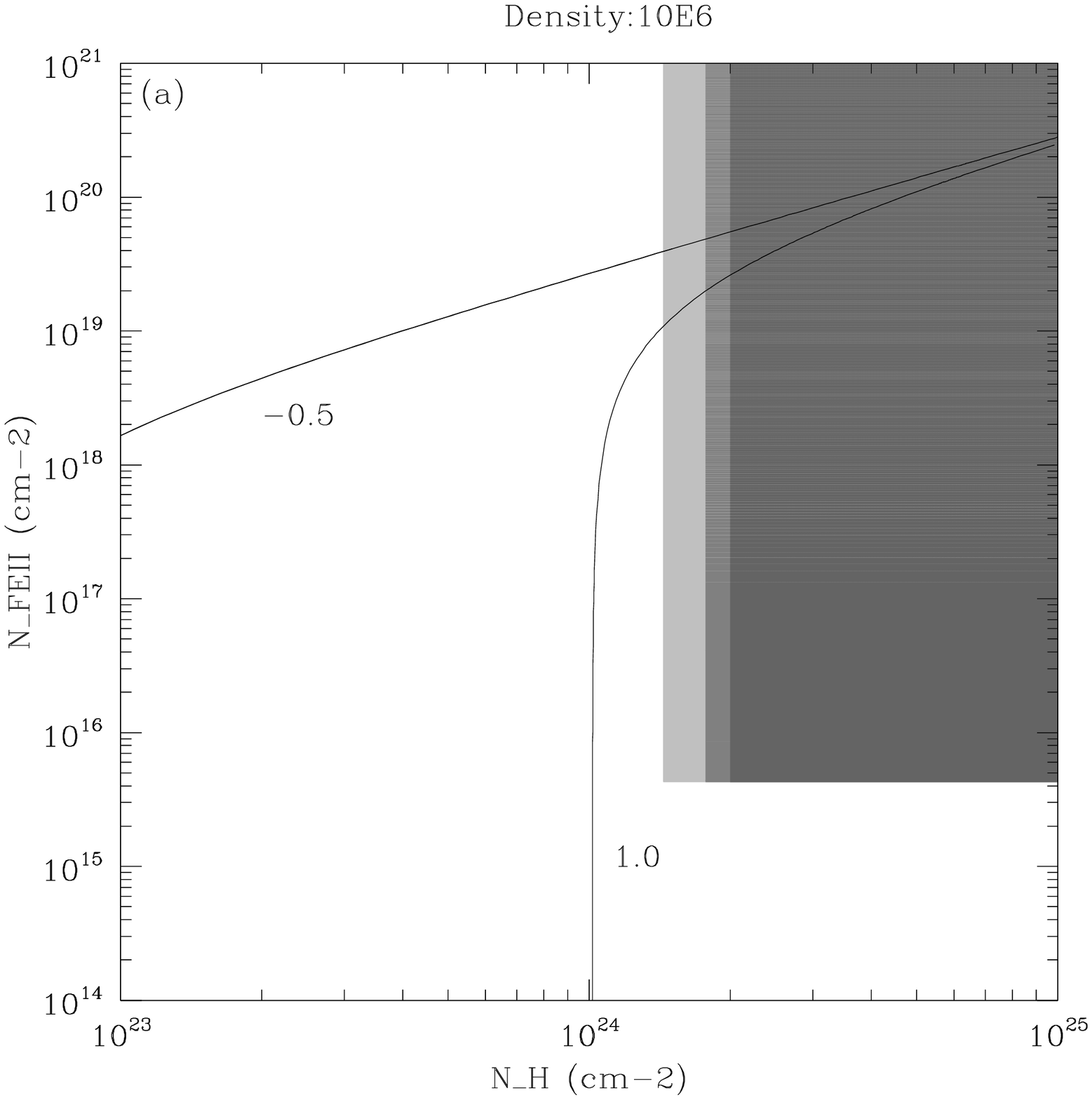}
\label{lowdensitycala}
}
\subfigure{
\includegraphics[scale=0.35]{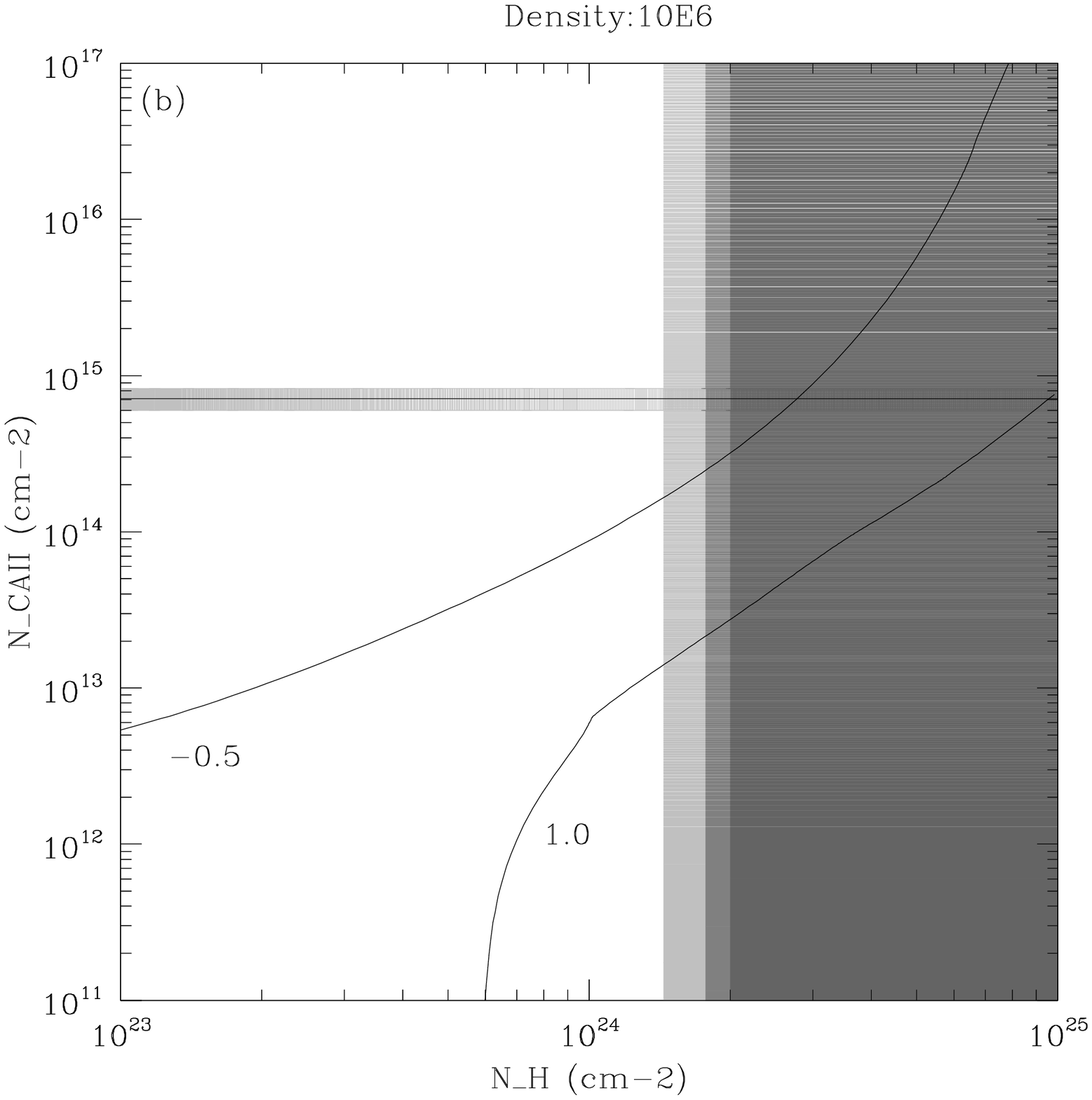}
\label{lowdensitycalb}
}

\caption{Two CLOUDY runs for $\log U=-0.5,1.0$ and $n_e = 10^{6}$ cm$^{-3}$.  The curves representing each ionization parameter are labeled.  The shaded areas represent the allowed column densities of {\bf (a)} \feii\ and {\bf (b)} \caii\ in \otbal\ based on UV observed column density limits and the lower limit, X-ray derived hydrogen column.}
\label{lowdensitycal}
\end{figure}

%---new n_e=1e6 plot goes here

%--------------------------END OF FIGURES-------------------------------------%
\clearpage
\appendix

\section{Fe\,{\sc iii} Column Density Lower Limits for SDSS J2215$-$0045}

The lower limit for the \feiii\ column density in \febal\ was
calculated (after normalizing by the unabsorbed continuum shown in Figure 1b)
using absorption from \feiii\ ions in the lower levels
of the UV34 and UV48 triplets (3.73 and 5.08 eV above ground, respectively).  
The individual components of each triplet are severely blended.  
Therefore, we treat each triplet's trough as coming from a single line with a 
$\lambda f$ value equal to the average of the components of the triplet, and 
then divide the resulting column density by three.  We assume that the covering
factor of the line is unity, so that the optical depth $\tau_{\msv}$ at each 
velocity \slv\ is given by $\tau_{\msv}=-\log l_{\msv}$, where $l_{\msv}$ is the absorbed 
continuum at $v$ normalized by the assumed unabsorbed continuum at $v$.  The 
resulting $\tau_v$ is always optically thin ($\tau_{\msv} \lesssim 0.6$).  We find
a raw lower limit of $\log N_{\rm FeIII}>15.18$ (lightest grey region in Figure
\ref{Iron3Graph}).
%by considering only \feiii\ in the lower levels of the UV34 and UV48 triplets 
%($\log N_{\rm FeIII}>15.03$ and $\log N_{\rm FeIII}> 14.68$, respectively).  
Even this is the largest \feiii\ column observed in any outflow to date
and is nearly ten times the $N_{\rm FeIII}$ observed in FBQS~0840+3633
by \cite{dek02b}.

We can set an improved $\log N_{\rm FeIII}$ lower limit 
by attempting to account for \feiii\ ions in other levels.  Assuming 
unsaturated troughs and that the excited \feiii\ level populations follow a 
Boltzmann distribution, the relative number densities in the two observed 
states yield $T_{ex}=34,250$\,K.  Accounting for \feiii\ ions in all levels 
with excitation energy $\leq 5.08$ eV at that temperature yields a factor of 28
correction from observed to total column in \feiii.  Thus the lower limit we 
adopt is $\log N_{\rm FeIII}>16.62$ (darker grey region in Figure 
\ref{Iron3Graph}).

It is important to note that the true \feiii\ column density 
could be substantially higher than any of the above estimates if
the troughs are saturated, which is quite likely \citep{sdss199}.
For example, M. Bautista (personal communication) finds a maximum possible 
ratio of $N_{\rm UV48}/N_{\rm UV34}=0.158$ at $n_e>10^{9.5}$\,cm$^{-3}$,
whereas we observe a value of 0.447 assuming unsaturated troughs.  
If we conservatively assume that only the UV48 trough is unsaturated
and use the maximum possible value of $N_{\rm UV48}/N_{\rm FeIII}=0.00155$,
again found at $n_e>10^{9.5}$\,cm$^{-3}$
(M. Bautista, personal communication), we obtain $\log N_{\rm FeIII}>17.49$. 
Note that 
%a density $n_e<10^{9.5}$\,cm$^{-3}$ would imply an even larger
%$N_{\rm FeIII}$;
a density $n_e=10^{7.5}$\,cm$^{-3}$ would imply $\log N_{\rm FeIII}=18.5$ and
a density $n_e=10^{6.5}$\,cm$^{-3}$ would imply $\log N_{\rm FeIII}=19.5$.

%\renewcommand{\theequation}{A-\arabic{equation}}
%\setcounter{equation}{0}  % reset counter 

%\input appendix.tex
%------------------------ END OF APPENDIX-------------------------------------%

\end{document}